\documentclass[fleqn,10pt]{wlscirep}
\usepackage[utf8]{inputenc}
\usepackage[T1]{fontenc}

\usepackage{svg}
\svgpath{{diagrams/}}
\usepackage{lineno}
\usepackage{float}
\usepackage{graphicx}
\usepackage{subcaption}
\usepackage{booktabs}
\usepackage{array}
\usepackage{threeparttable}
\usepackage{tabularx}
\usepackage{subcaption}
\title{Two-shot learning of multiple strange attractors}

\author[1,*]{Daniel K{\"o}glmayr}
\author[2,*]{Miralem Spahic}
\author[3,4]{Andrew Flynn}
\author[5,2]{Christoph R{\"a}th}
\affil[1]{Institut f{\"u}r KI-Sicherheit,
Deutsches Zentrum f{\"u}r Luft- und Raumfahrt (DLR), Ulm, Germany}
\affil[2]{Ludwig-Maximilians-Universit{\"a}t (LMU), M{\"u}nchen, Germany}
\affil[3]{School of Mathematical Sciences, University College Cork, T12 XF62 Cork, Ireland}
\affil[4]{INFANT Research Centre, University College Cork, T12 DC4A Cork, Ireland}
\affil[5]{Institut für Frontier Materials auf der Erde und im Weltraum, Deutsches Zentrum f{\"u}r Luft- und Raumfahrt (DLR), 51140 K\"oln, Germany}
\affil[*]{These authors contributed equally to this work. Corresponding author: daniel.koeglmayr@dlr.de}

\begin{abstract}

The brain combines short- and long-term memory to process, store, and recall multiple different pieces of information. Inspired by this and recent results on multifunctional and parameter-aware learning, we extend a new machine learning technique that combines short- and long-term memory units, specifically, a system consisting of a next-generation reservoir computer (NGRC) and extremely randomized trees (ERT), to process, store, and recall multiple different strange attractors.
We train the combined NGRC+ERT system using a two-shot learning approach which significantly improves performance by filtering out unnecessary features, thereby avoiding extensive hyperparameter optimization.
We first show that an NGRC+ERT system achieves highly accurate reconstruction of the short- and long-term dynamics of both the Lorenz and Halvorsen chaotic attractors when using an exponential filtering scheme. We validate these finding by training the NGRC+ERT system to reconstruct different pairs of attractors and also a greater number of attractors. 
We focus on the task of training a single NGRC+ERT system to reconstruct 16 different attractors and show that sufficient index-based separation in feature space suppresses unwanted switching dynamics, thus stabilizing long-term memory recall. Finally, we identify that defects in short-term memory processing can provoke failure modes in long-term memory recall resulting in confabulation. 

\end{abstract}

\usepackage{xr-hyper}
\begin{document}

\flushbottom
\maketitle

\thispagestyle{empty}

\section*{Introduction}

% The ability of the brain to combine short- and long-term memory to rapidly process, store, and recall complex temporal patterns without extensive retraining represents a fundamental computational principle that has long inspired research in artificial intelligence. 

The brain has been a constant source of inspiration for advancing and understanding the capabilities of artificially intelligent systems. The mathematics of dynamical systems theory often acts as a bridge between these worlds where developments in one area can inform the other\cite{raut2025arousal}.
In this paper we take inspiration from the brain's ability to combine short- and long-term memory to rapidly process, store, and recall multiple complex temporal patterns.
%{\af\st{ without extensive retraining}}.
To do this, certain biological neural networks use `multifunctionality', when a single network exploits its capacity for multistable dynamics to perform multiple different tasks\cite{getting1989emerging,dickinson1995interactions,marder1996principles}. 
% However, how these single neural networks process, store, and then recall multiple functionalities remains far from understood. 
% In parallel, the application of dynamical systems in neuroscience has gained significant attention. For instance, 
%{\af\st{Furthermore, low-dimensional time-delay embeddings of a scalar arousal index (pupil diameter) were shown to be used by the brain to reconstruct high-dimensional spatiotemporal neural dynamics.}\em(nice but out of place)}
% , supporting the hypothesis that arousal operates as a low-dimensional dynamical system coordinating brain-wide activity \cite{raut2025arousal}.
%{\st{Researchers showed that the brain's arousal system operates as a low-dimensional dynamical system that orchestrates brain-wide activity, whereby time-delay embeddings of the pupil diameter time series contains sufficient information to reconstruct high-dimensional spatio-temporal brain states .}}
% {\af\em(the opening paragraph do not gel well, needs more work)}
%{\af\st{Learning, predicting and inferring aspects of dynamical systems is an active area of research. }}
With advances in machine learning, researchers are increasingly focused on developing computational models that capture key characteristics of biological neural networks, including their ability to process, store and recall temporal information or adapt to changing inputs\cite{kim2021teaching,yamazaki2022spiking}.
%{\dk cite more here, maybe also SNN ?}.
These efforts bridge 
% {dynamical systems theory,}
neuroscience and modern machine learning, offering new tools for modeling neural computation. 
In particular, recent studies demonstrate that multifunctional reservoir computers (RCs)\cite{herteux2020breakingsymmetries,flynn2021multifunctionality,flynn2021symmetry,flynn2022exploring,kong2024reservoir,terajima2025multifunctional}, a class of recurrent neural networks, provide an effective framework for storing and retrieving temporal information from multiple dynamical systems within a single network. 
While traditional networks, such as Hopfield networks are limited to storing static patterns, RCs can process, store, and recall multiple strange attractors through either index-based (location-addressable/parameter-aware) or multistability-based (content-addressable/multifunctional) approaches \cite{kong2024reservoir}. Building on these memory storage capabilities, researchers have began investigating the associated failure modes and limitations of RCs\cite{flynn2025confabulation}.
%{\af\st{For instance, O'Hagan \textit{et al.}~ recently investigated the role played by complete confabulations in the form of `untrained attractors' in attractor reconstruction tasks. Theses are attractors that were not present during training but the RC generates to fill gaps in its memory, a phenomenon akin to a false memory and first observed by Flynn \textit{et al.}~ in a multifunctional RC. Further, another type of confabulation, a `momentary confabulation' in the form of a `generated attractor' was studied by O'Hagan \textit{et al.}~ where the prediction only partially resembles the original attractor and is generated from deformations of the attractor.}}
% {\af\st{The study demonstrates that complete confabulations persist even with additional training data, suggesting that such false memories are a fundamental characteristic of bounded learning systems, potentially including the brain and modern AI. }\em(I think this might be too much detail for the present paper.)}
% {\af\st{However, a mechanistic explanation for why these confabulations emerge remains unclear.}\em(not true, explained in my paper.)}

% {\af\st{Achieving multifunctionality in reservoir systems} }
Training a single RC to reconstruct more than one attractor comes with significant challenges.
In the case of multifunctionality, optimizing the RC requires extensive hyperparameter tuning, particularly when reconstructing overlapping attractors~\cite{flynn2022exploring,flynn2023seeing}.
Classical RCs use high-dimensional recurrent networks with fixed weights and a trainable readout layer, typically trained using ridge regression. Recent work proved the mathematical equivalence of nonlinear autoregressive models and RCs \cite{bollt2021explaining}, inspiring next-generation reservoir computing which replaces recurrent networks with explicit monomial expansions of time-delayed states, offering reduced computational cost, fewer hyperparameters, and improved interpretability \cite{gauthier2021next}. 
More recently, Giammarese et al.~\cite{giammarese2024tree} demonstrated that the output layer of a next-generation reservoir computer (NGRC) can be successfully trained using Extremely Randomized Trees (ERT), a tree-based ensemble method, rather than conventional ridge regression ans show their approach exhibits improved robustness and high-fidelity prediction of chaotic attractors. Crucially, the tree-based approach enables extraction of feature importance measures directly from the trained trees, allowing for substantial improvements in hyperparameter tuning by retraining with features identified as relevant for the specific task.

In this work, we take inspiration from the results on multifunctional and parameter-aware reservoir computing described above to extend the capabilities of the NGRC+ERT system and enable it to process, store, and recall multiple complex temporal patterns. 
% {\af\st{the combination of NGRC and ERT to investigate multifunctionality for memorizing multiple dynamical attractors.}} 
% {\af
% \st{Drawing inspiration from the brain as well, we model the interplay between neural regions through a clear separation of short-term and long-term memory processing. Specifically, we utilize time-delay embeddings derived from NGRC as the short-term memory unit and ERT as the long-term memory unit, enabling a single NGRC+ERT system to process, store, and recall multiple different strange attractors.} 
Further, in the brain, short-term memory is primarily associated with the prefrontal cortex, long-term memory with the cerebral cortex, and the hippocampus serves as the critical bridge between these regions to enable memory consolidation\cite{WhereAreMemroiesStored,tulving1995organization}.
In our case, the NGRC acts a short-term memory unit, the ERT as a long-term memory unit, and by combining these we produce a NGRC+ERT system capable of achieving accurate short-term predictions and long-term reconstruction of multiple strange attractors.
% without retraining \dk whats retraining here ? We train twice.}
% More specifically, we extend and optimize the work of Giammarese et al.~\cite{giammarese2024tree} for 
% {\af\st{multifunctionality}
% multi-attractor learning and
We introduce a two-shot learning methodology 
% {\af\st{for learning multiple dynamical attractors}
which significantly improves performance and reduces the need of extensive hyperparameter tuning. 
% {\af\st{Our methodology proceeds in two steps. First,} 
For the first-shot we use NGRC hyperparameter configurations to generate an initial feature representation of the input data and train an ERT model. 
% {\af\st{Subsequently,}
For the second-shot we employ 
% {\af\st{the tree ensemble's}}
feature importance metrics to identify important features, 
% {\af\st{ through an exponential feature importance criterion. This enables a refinement of the feature representation to obtain an optimized}
which notably amplify short-term memory processing 
% {\af\st{configuration used} 
features,
% }
to train a 
% {\af\st{second and final ERT instance}
new NGRC+ERT system which significantly outperforms the initial system.
% }. {\af\st{We call this approach two-shot learning for temporal dynamics.}} 
To scale this approach for learning multiple dynamical attractors within a single 
% {\af\st{architecture} 
system,
% },
we use an index-based (location-addressable) approach 
% {\af
which allows the NGRC+ERT system to operate in either a multifunctional or parameter-aware style setup.
% }. 
During training, each attractor's feature representation receives a unique identifier that separates the different attractor dynamics in feature space. When the system recalls a specific attractor, it is because its corresponding identifier has been activated.
% {\af\st{This} 

Our overall
% }
approach offers two key advantages over current methods. First, it enables interpretable separation of short-term and long-term memory properties, with NGRC handling the short-term memory processing through time-delay features while ERT exclusively provides the long-term memory recall of the temporal dynamics. Second, the approach allows an optimization of 
% {\af\st{short-term}}
memory properties using the tree-based feature importance measures, making extensive hyperparameter optimization superfluous.
We develop and test the approach on the well-studied overlapping Lorenz and Halvorsen task\cite{herteux2020breakingsymmetries,flynn2022exploring,flynn2025confabulation} and validate our findings on similar Chua and Halvorsen, and Rucklidge and Windmi tasks.
When applied to a 16-overlapping-attractor task, we find that without sufficient feature separation the NGRC+ERT system suffers from unwanted switching dynamics like in Flynn and Amann\cite{flynn2024switching} and Kabayama \textit{et al.} \cite{kabayama2025crisis}, indicating long-term memory recall problems. 
% {\af\st{However, with sufficient index-based feature separation successfully suppresses the switching dynamics, demonstrating the NGRC+ERT system's capability to maintain stable long-term memory recall over extended periods for large amounts of attractors. }\em(repetition)}
Consequently, the interpretable nature of short-term memory processing through time-delay features allows us to analyze how defects in short-term memory, resembling deterioration of the system in `on-chip' applications, trigger long-term memory recall errors, causing the NGRC+ERT system to generate 
% {\af\st{confabulating predictions of attractors it was never trained on}
confabulations in the form of `generated attractors' — a malfunction that may represent a fundamental characteristic shared across various learning systems\cite{flynn2025confabulation}.
% , potentially including biological and modern artificial neural networks.

\section*{Results}

\subsection*{Two attractor task}

\begin{figure}[t]
    \centering
    \includegraphics[width=\linewidth]{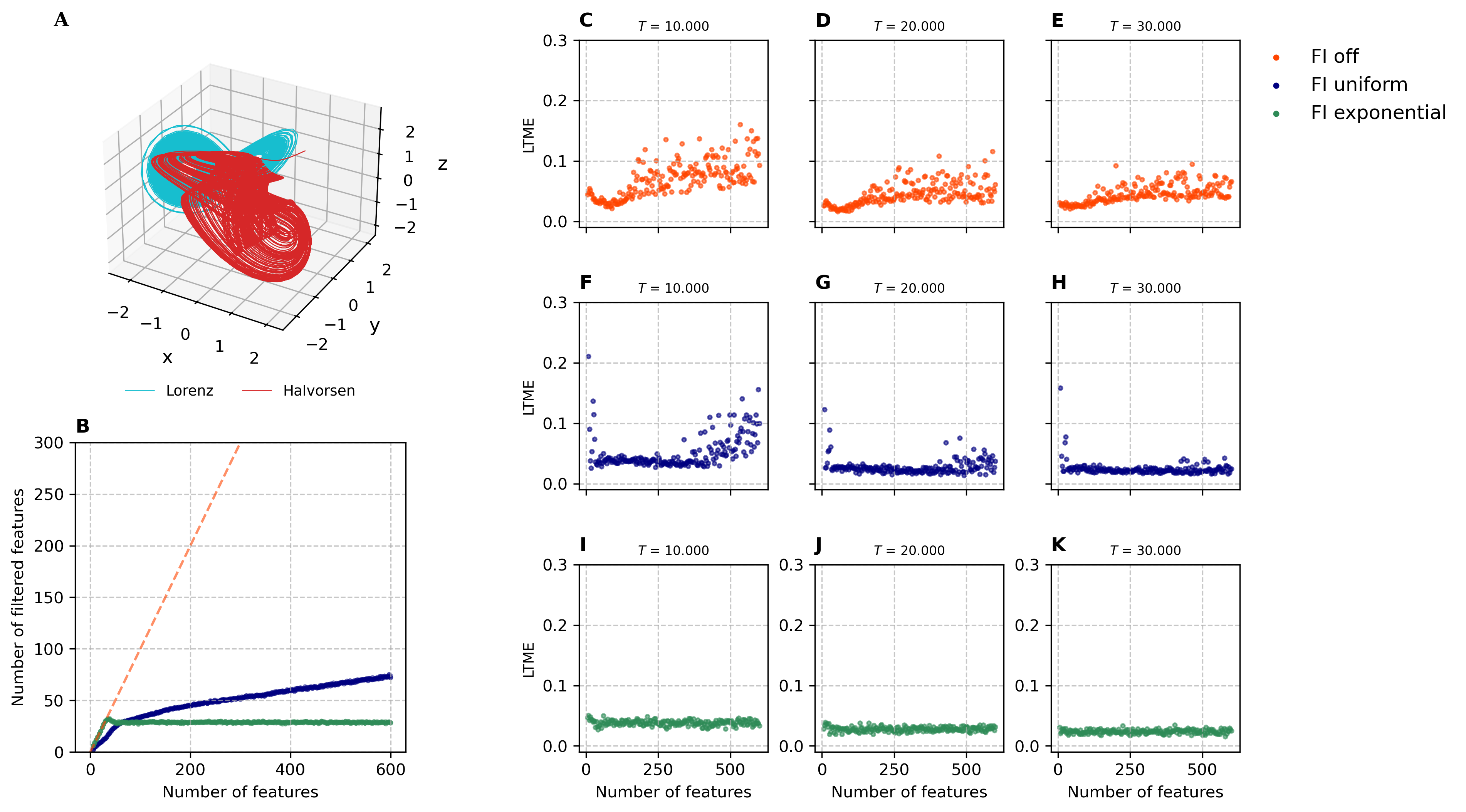}
    \caption{\textbf{Feature importance filtering improves attractor reconstruction in VAR feature model for the Lorenz and Halvorsen task.} \textbf{A} Overlapping trajectories of the Lorenz and Halvorsen attractors after normalization. \textbf{B} Comparison of filtered versus initial features for three approaches: No filtering (base model) (top), uniform filtering (middle), and exponential filtering (bottom) for measured feature importance FI. 
    % {\af\st{The uniform filtering shows linear growth in retained features, while the exponential filtering provides around ~28 retained features beyond 40 initial input features.} \em(no detailed discussion of figures in caption)}
    \textbf{C}-\textbf{E} Long-term measurement error for  VAR hyperparameter configurations without FI filtering. 
    % {\af\st{Error minimizes with few features and increases with additional features, with this degradation less pronounced for larger training sizes, indicating an optimal feature configuration.}}
    \textbf{F}-\textbf{G} Long-term measurement error using uniform filtering. 
    % {\af\st{Performance matches the base model minimum across a broad parameter range, though error increases with larger numbers of features for $T=10{,}000$, indicating retention of irrelevant features.}}
    \textbf{I}-\textbf{K} Long-term measurement error (LTME) using exponential filtering. 
    % {\af\st{Error remains stable and comparable to base model minimum across all configurations while using substantially fewer and constant amounts of features.}} 
    Each data point represents the mean over ten random initializations of ERT instances.}
    \label{fig:Lorenz + Halvorsen}
\end{figure}

This subsection presents results for the Lorenz and Halvorsen task (see Methods). We train an NGRC+ERT system to reconstruct these two chaotic attractors that overlap in state space %{\ms state or phase space? In figures/supplementary phase space was used, check for consistency}{\af state space is technically correct, we are looking at the space where states exist not the space where phases exist}
(Figure~\ref{fig:Lorenz + Halvorsen}~\textbf{A}) using linear time-delayed features (VAR feature model, see Methods). The number of features created for each configuration is $N_f = k \times d$, where $k$ is a time-delay parameter used to denote the number of time-delayed data points and $d = 3$ represents the dimension of the corresponding Lorenz and Halvorsen systems.
%the dimensionality of the attractors we work with in this paper are fractal, between 2 and 3.
% {\af\st{For the `VAR' feature model, we varied the parameter} 
We evaluated the performance of the NGRC+ERT system in this setup for different choices of $k$ from $2$ to $200$, while keeping the time steps between the time-delayed data points fixed at $s = 1$. 
% {\af\st{Learning and predicting overlapping attractors poses a significant challenges, as the partial occupation of shared spatial regions prevents the model from easily distinguishing between the dynamics on each attractor. }\em(ONLY a problem for multifunctionality. The index method below is not multifunctionality, it is parameter-aware in disguise. relationship between parameter-aware performance and overlapping attractors has not been studied extensively, I did it in task 2 and 3 of my confabulation paper.)\dk fine for me! }
Here, we employ an index-based separation strategy to distinguish between attractors in feature space. Specifically, we add a scaled constant $\theta_0 = -1$ times scaling parameter $\gamma = 1$ to each feature derived from the Lorenz data, likewise $\theta_1 = 1$ times $\gamma = 1$ to the features derived from the Halvorsen data.
This modification enables the NGRC+ERT system to differentiate between the two attractors during both training and prediction; the NGRC+ERT system reconstructs a different attractor depending on the choice of $\theta$.
We assess attractor reconstruction accuracy using metrics that quantify short- and long-term prediction behavior, as detailed in the Methods section.
%behavior of the predictions by comparing the largest Lyapunov exponent \cite{rosenstein1993practical} (LY) and correlation dimension \cite{grassberger2004measuring} (CD) of reconstructed and original attractors over 15{,}000 time steps, by means of attractor reconstruction \cite{lu2018attractor}.
%This comparison is quantified by the average relative error between the LY and CD for each recalled and original attractor in the two attractor task. A small error indicates that the NGRC+ERT system successfully achieves multi-attractor reconstruction, accurately recalling both attractors and maintaining precise long-term memory recall of each over extended periods. We name this measure for the two attractor tasks long-term measurement error.
Results on short-term prediction behavior are provided in Fig. S2 of the Supplementary Information, along with supplementary analysis for the NVAR and NML feature models %{\af in Figs.~X,Y ... please reference s1-s4 somewhere if possible and all figs which stem from points in the main so they dont just `appear out of nowhere' in the supp}.

%{\af\em(The concept of small $k$ and short-term memory appear to be equated here, I found this highly confusing. It would be better to discuss $k$ in terms of what it is, a delay term.) old:\dk clarified ?\em(yes needs to be clarified)}
Figures~\ref{fig:Lorenz + Halvorsen} ($\mathbf{C}$–$\mathbf{E}$) show the long-term measurement error (LTME) of the reconstructed attractors vs. $N_f$ for different training sizes $T$.
We observe a minimum in the error metric for a similar $N_f$ across all training sizes, indicating that there is an optimal choice of $k$ that enables the NGRC+ERT system to perform accurate long-term reconstruction of each attractor. Notably, this performance deteriorates significantly when $k$ is too large but can be improved by using larger $T$.
% {\af\st{, impairing the ability to maintain long-term memory recall over extended periods}}.

We evaluate the feature importance (FI) measure inherent to the ERT (see Methods) to optimize the feature configuration. 
% {\af\st{Assessing the feature importance of the initially provided features after training enables a filtering strategy that removes features identified as irrelevant. We retrain the NGRC+ERT system while retaining only those features whose importance exceeds the threshold $\mathrm{I}_c$.}
In short, we train a new NGRC+ERT system with features whose importance exceeds the threshold $\mathrm{I}_c$.
We employ the `uniform filtering scheme' introduced in the TreeDOX framework\cite{giammarese2024tree}, in which the threshold is defined as $\mathrm{I}_c = 1/N_f$. Figures~\ref{fig:Lorenz + Halvorsen} ($\mathbf{F}-\mathbf{H}$) show the resultant LTME from uniform filtering. We find that long-term reconstruction improves in all analyzed training sizes, with the improvement being particularly prominent for larger training sizes. Comparing Figure~\ref{fig:Lorenz + Halvorsen} $(\mathbf{C}) ~\text{to}~ (\mathbf{F})$, we observe that uniform filtering broadens the range of $N_f$ values where similar long-term measurement errors are obtained, specifically $N_f$ between approximately $60$ and $350$. For small values of $N_f$, the uniform filtering approach results in higher errors, while for $N_f > 350$ no notable improvement is observed. Results on short-term prediction behavior and supplementary analysis for the NVAR and NML feature models are provided in Fig.~S3 of the Supplementary Information.
% {\af\st{By analyzing the proportion of features retained after filtering relative to the initially provided number of features, we observe that uniform filtering leads to a nearly linear increase in the number of retained features for $N_f > 100$ (Figure~\ref{fig:Lorenz + Halvorsen} $\mathbf{B}$, blue). The increase in error for $N_f > 350$ is explained by comparing the number of retained features with the optimal number of feature observed in Figs.~\ref{fig:Lorenz + Halvorsen} ($\mathbf{C}$–$\mathbf{E}$). For large $N_f$ , the uniform filtering approach classifies too many features as important, thereby degrading predictive performance.}}

To counteract the error distribution observed, we introduce an `exponential filtering scheme'. We set the cut-off value to the mean lifetime of exponential decay, calculated as $\mathrm{I}_c = \mathrm{I}_{max}/e$, where $\mathrm{I}_{max}$ is the maximum measured feature importance value. 
In Figs.~\ref{fig:Lorenz + Halvorsen} ($\mathbf{I}-\mathbf{K}$) we plot the resultant LTME from exponential filtering versus $N_f$. These show that the exponential filtering scheme provides low and constant LTME for each $N_f$ and that increasing $T$ results in marginally smaller LTME.
% {\af\st{The LTME using exponential filtering are shown in Figs.~\ref{fig:Lorenz + Halvorsen} ($\mathbf{I}-\mathbf{K}$).}}

Furthermore, our results reveal distinct differences between the uniform and exponential filtering schemes. For instance, 
Fig.~\ref{fig:Lorenz + Halvorsen} $\mathbf{B}$ shows the number of filtered features versus the number of features that are used to initially train the NGRC+ERT system. For uniform filtering (blue curve) we observe an almost linear relationship for $N_f>100$. In contrast, exponential filtering (green curve) converges toward a constant number as $N_f$ increases. Further, for small $N_f$, exponential filtering performs no feature reduction, whereas uniform filtering does, i.e., deviates from the orange curve. 
Based on these results, the increase in LTME for $N_f>350$ in Figs.~\ref{fig:Lorenz + Halvorsen} ($\mathbf{C}$–$\mathbf{E}$) may be correlated with the number of filtered features increasing past a certain point in the uniform filter case.
Supplementary Figures~S4--S6 show the short-term prediction behavior, reconstructed attractors obtained with exponential filtering, and relative improvements achieved by both filtering schemes compared to results without filtering, respectively. The exponential filtering scheme achieves LTMEs across all $N_f$ similar to the minimal LTME obtained without filtering, thereby eliminating the need for extensive hyperparameter optimization.
%{\af\em(what are fig S1-S4 related to in the main text?)}
%{\dk \st{We show here that the long-term {\af{memory recall}\em(what is this?)} achieved with exponential filtering across all initial number of features is comparable to the optimum obtained without filtering, thereby eliminating the need for extensive hyperparameter optimization.}}

% {\af \em(integrated with text above) Fig.~\ref{fig:Lorenz + Halvorsen} $\mathbf{B}$ shows that while uniform filtering exhibits a nearly linear increase in the number of retained features, exponential filtering converges toward a constant number as $N_f$ increases (Figure~\ref{fig:Lorenz + Halvorsen} $\mathbf{B}$, green). For small $N_f$, exponential filtering performs no feature reduction, whereas uniform filtering does. Supplementary Figure S6 shows the relative improvements achieved by both filtering methods compared to the results obtained without filtering. The long-term memory recall achieved with exponential filtering across all initial number of features is comparable to the optimum obtained without filtering, thereby eliminating the need for extensive hyperparameter optimization. Supplementary Figure S5 shows the predicted attractors obtained with exponential filtering applied across different hyperparameter configurations and training sizes.
% }

We validate our findings by showing that similar behavior is observed across 
% across {\af{initial}\em(what is initial here? no filtering?)\dk ja no filtering so whats used in the first shot, but we can just leave it, or further specify that the same NGRC hyperparameter were used as for this task}
hyperparameter configurations before filtering when using different attractors as training data, specifically, we analyze the proposed exponential filter scheme on overlapping Chua and Halvorsen attractors as well as on an overlapping Rucklidge and Windmi attractors in the Supplementary Information. 
% {\dk \em commented out the lower abstract as this is reduent with this abstract here, and further discussed in the discussion.}

%{\em (For each task, exponential filtering identifies and maintains an {\af{performant}\em(what is this?) \dk low and constant LTMEs across hyperparameter configurations} {\af{short-term memory} \em(I'm guessing you mean small $k$?)} configuration, enabling long-term memory recall across {\af{initial}\em(what is meant by initial) \dk same initial here, can be canceled} hyperparameter configurations.) }

\subsection*{Multiple attractor task}

\begin{figure}[t]
    \centering
    \includegraphics[width=0.9\linewidth]{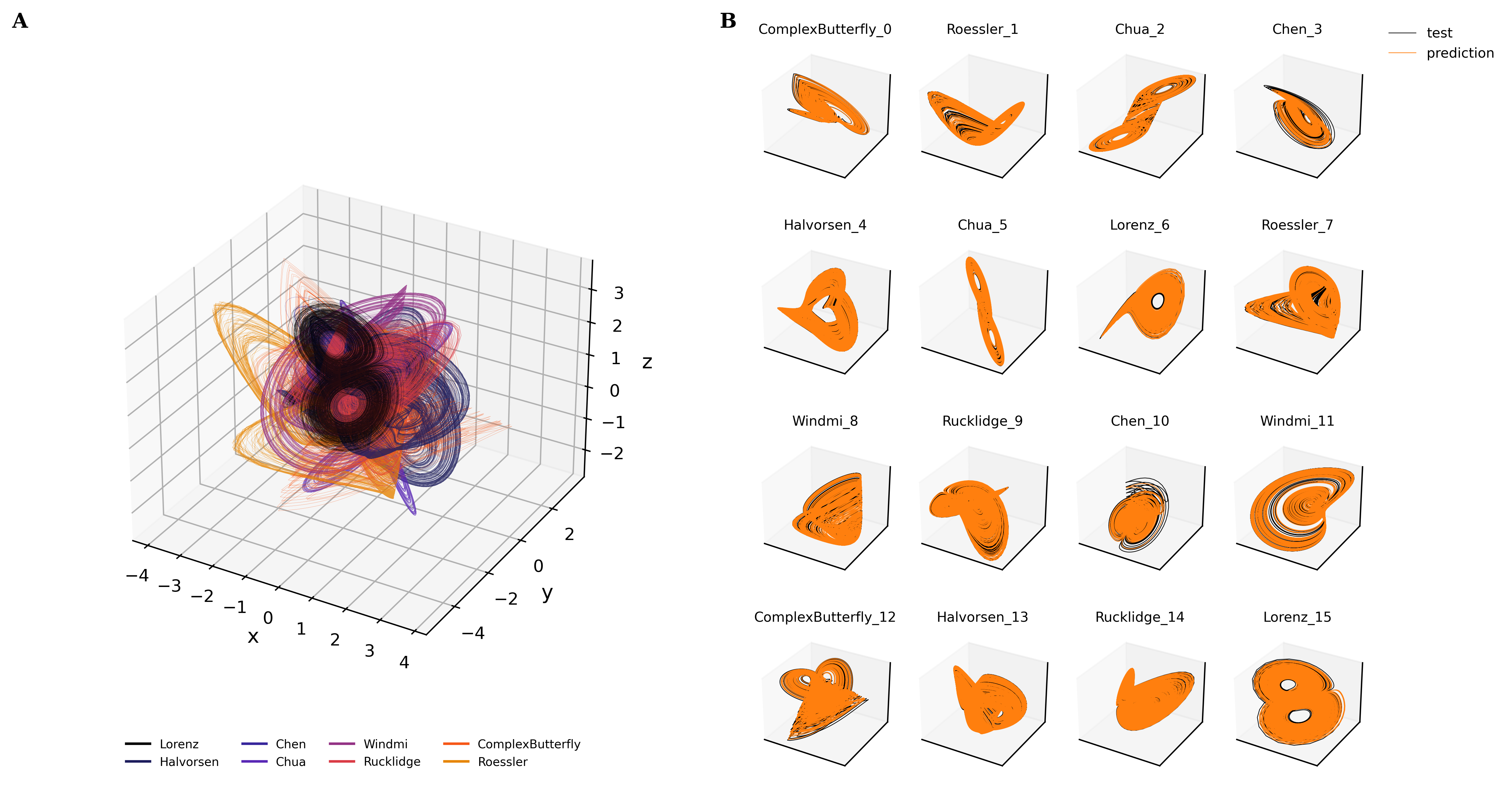}
    \caption{\textbf{Multi-attractor reconstruction}. \textbf{A} Experimental setup defined by 16 overlapping attractors in state space, each normalized to zero mean, unit standard deviation and random rotation. \textbf{B} NGRC+ERT predictions with exponential filtering applied.}
    \label{fig:prediction of 16 attractor for high scaling parameter}
\end{figure}

This subsection presents results from training an NGRC+ERT system to reconstruct more than two attractors. Specifically, we train the NGRC+ERT system on the 16-overlapping-attractor task specified in the Methods section, see Fig.~\ref{fig:prediction of 16 attractor for high scaling parameter} $\mathbf{A}$ for a plot of these attractors overlapping in the same space. Informed by Fig.~\ref{fig:Lorenz + Halvorsen}, we keep the training parameters as $T=20{,}000$, $k=100$, $s=1$, choose $\gamma=25$ and assign a unique identifier $\theta \in \{-8,\,-7,\,\dots,\,7,\,8\}\setminus\{0\}$ to each attractor. We train the NGRC+ERT system using the same two-shot learning approach with the exponential filtering scheme. 
We assess its performance by computing the forecast horizon, largest Lyapunov exponent (LY), and correlation dimension (CD) of each reconstructed attractor.
We compute the mean and standard deviation of these metrics for each attractor based on the trajectories obtained from 100 different random initializations of the NGRC+ERT system and 100 different realizations of the ERT, i.e., based on 10,000 calculations.
% we do this for 100 
% taking the mean and 
% To account for the randomness involved in ERT, we compute the performance metrics by computing the mean and standard deviation across 100 different random initializations.
% Further, for each NGRC+ERT system we train

%{\af{More specifically, we train the NGRC+ERT system for 100 different random initializations of the ERT and compute the mean and standard deviations of the performance metrics described below. Further, in each instance, we compute the performance metrics based on 100 different random initialisations of the NGRC+ERT system. In total we perform 10,000 }\em(needs more work)}
% \st{, by first optimizing the short-term memory by an initially trained NGRC+ERT system and train the system a second time with optimized features.}}
% {\af\st{To account for the randomness involved in the NGRC+ERT system, we quantify the short- and long-term behavior of the attractors reconstructions with 100 different random initializations of the ERT.}}

% {\af\st{We show in the }
In Table~\ref{tab:16} we list the attractors we consider and the integration time step used to generate a trajectory on the attractor in the first and second columns, the mean forecast horizon, LY, and CD of the reconstructed attractor, together with their standard deviations, in the third, fourth, and fifth column and compare these results to the corresponding LY and CD of the original attractors in the sixth and seventh columns. 
NGRC+ERT with exponential filtering achieves a mean forecast horizon of 3.39 Lyapunov times, a LY mean absolute error of 0.041, and a CD mean absolute error of 0.019 averaged across all attractor reconstructions, demonstrating strong short-term prediction accuracy and successful long-term reconstruction of the chosen attractors. 

\begin{figure}[H]
    \centering
    \includegraphics[width=\linewidth]{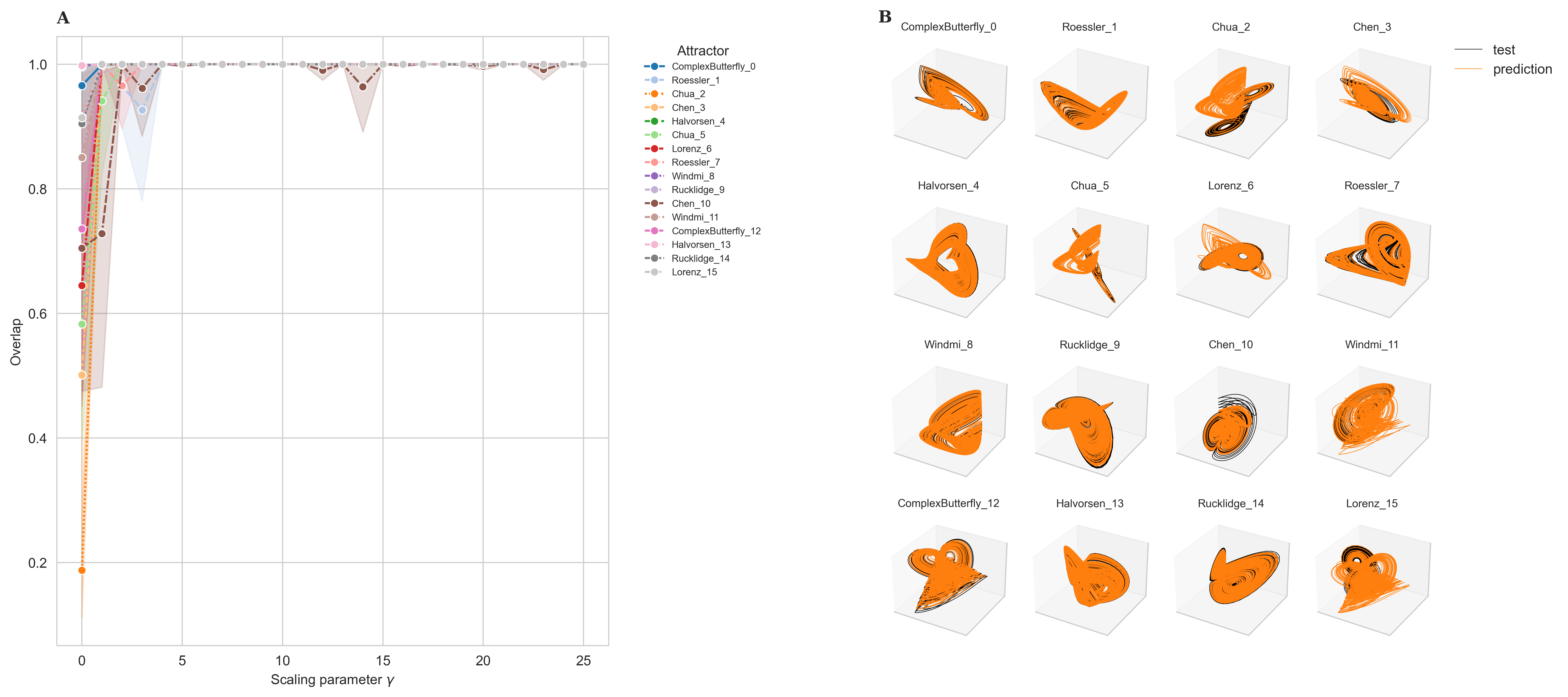}
    \caption{\textbf{Effect of scaling parameter $\gamma$ on attractor reconstruction.} \textbf{A} Overlap metric as a function of $\gamma$ for all 16 attractors in the multi-attractor reconstruction setup. 
    % {\af\st{Overlap quantifies the ratio of leaves associated with predicted versus training outputs for each trajectory. Values below 1 indicate misidentification of trajectory segments as belonging to different attractor dynamics.}}
    \textbf{B} Predicted trajectories for 16 attractors with $\gamma = 0$ (content-addressable approach), showing spontaneous switching between trained attractors in panels titled `Chua\_2', `Chen\_3', `Chua\_5', `Lorenz\_6', and `Lorenz\_15'. 
    % {\af\st{Increasing $\gamma$ suppresses switching frequency by enhancing feature space separation.}}
    }
    \label{fig:spontaneous switching}
\end{figure}

We find that by choosing the scaling parameter, $\gamma$, to be small, predictions become susceptible to failure modes such as switching dynamics or recalling attractors from different $\theta$.
Figure~\ref{fig:spontaneous switching} $\mathbf{B}$ shows examples of this switching behavior for a purely content-addressable approach (implemented by setting $\gamma=0$); see panels titled `Chua\_2', `Chen\_3', `Chua\_5', `Lorenz\_6', and `Lorenz\_15'.
% {\af \st{They} 
These switching dynamics resemble an attractor consisting of two quasi-attracting regions that correspond to two attractors that may have previously coexisted for a different choice of hyperparameters. We note that similar phenomena has been found when training traditional RCs to achieve multifunctionality; see Flynn and Amann~\cite{flynn2024switching} and Kabayama \textit{et al.}~\cite{kabayama2025crisis} for further details. 
% {\af\st{Figure~\ref{fig:spontaneous switching} $\mathbf{B}$ shows the switching behavior for a purely content-addressable approach (implemented by setting $\gamma=0$). Thereby, the NGRC+ERT system cannot maintain the initial short-term memory assignment, preventing proper recall of the associated long-term memory.}} 

To obtain further insight into these switching dynamics for specific $\gamma$, we analyze which output leaves of the ERT are traversed during prediction of a given attractor.
% , i.e., we compare the leaves traversed during switching dynamics to the leaves traversed during training.
Specifically, we define an overlap measure as the ratio of leaves active during prediction to those associated with each unique identifier $\theta$ during training. An overlap smaller than 1 indicates that parts of the predicted trajectory follow different dynamics to those used during training, which we classify as switching dynamics. We quantify this behavior for scaling parameters $\gamma$ ranging from 0 to 25. 
In Fig. \ref{fig:spontaneous switching}~$\mathbf{A}$, we plot the corresponding overlap measure for each attractor across different scaling parameters, taking into account the inherent randomness of ERT by performing the analysis over 10 random initializations. The results show that larger scaling parameters reduce the frequency of switching behavior, with some large scaling parameters completely suppressing the switching behavior across all predicted attractors. 
Large scaling parameters, and hence larger separation of dynamics in feature space, provide a mechanism to stabilize predictions by controlling the susceptibility of the ERT unit to switching between leaves assigned to different attractors during prediction. This index-based approach effectively suppresses the switching behavior and enables stable long-term memory recall. Although we do not claim that this method achieves the highest precision forecasts in systems trained to perform multi-attractor reconstruction, our results demonstrate that this index-based approach enables the NGRC+ERT system to successfully reconstruct the dynamics of multiple attractors without extensive hyperparameter optimization.

\begin{table}[H]
    \centering
    \resizebox{0.95\linewidth}{!}{%
    \begin{tabular}{ll|ccc|cc}
        \toprule
        & \multicolumn{4}{c|}{Predictions} & \multicolumn{2}{c}{Test Data} \\
        \cmidrule(lr){3-5} \cmidrule(lr){6-7}
        Attractor & $\Delta$ t & Forecast Horizon & Larg. Lyapunov Exp. (LY) & Corr. Dim. (CD) & LY & CD \\
        \midrule
        $C. Butterfly\_0$   & 0.08      & 5.07      $\pm$ 0.59          & 0.13       $\pm$ 0.01      & 2.19     $\pm$ 0.04    & 0.17             $\pm$ 0.0             & 2.25           $\pm$ 0.01          \\
        $Roessler\_1$           & 0.1       & 4.65      $\pm$ 0.45          & 0.08       $\pm$ 0.01      & 1.88     $\pm$ 0.07    & 0.08             $\pm$ 0.0             & 1.85           $\pm$ 0.01          \\
        $Chua\_2$               & 0.02      & 2.55      $\pm$ 0.37          & 0.3        $\pm$ 0.03      & 1.98     $\pm$ 0.07    & 0.35             $\pm$ 0.01            & 1.93           $\pm$ 0.01          \\
        $Chen\_3$               & 0.005     & 2.98      $\pm$ 0.74          & 2.02       $\pm$ 0.18      & 2.12     $\pm$ 0.03    & 1.94             $\pm$ 0.1             & 2.12           $\pm$ 0.01          \\
        $Halvorsen\_4$          & 0.02      & 1.61      $\pm$ 0.07          & 0.69       $\pm$ 0.04      & 2.11     $\pm$ 0.0     & 0.79             $\pm$ 0.02            & 2.11           $\pm$ 0.0           \\
        $Chua\_5$               & 0.02      & 2.65      $\pm$ 0.43          & 0.3        $\pm$ 0.04      & 1.96     $\pm$ 0.05    & 0.35             $\pm$ 0.01            & 1.93           $\pm$ 0.01          \\
        $Lorenz\_6$             & 0.02      & 5.41      $\pm$ 1.25          & 0.85       $\pm$ 0.03      & 2.06     $\pm$ 0.01    & 0.84             $\pm$ 0.01            & 2.06           $\pm$ 0.01          \\
        $Roessler\_7$           & 0.1       & 4.04      $\pm$ 0.38          & 0.07       $\pm$ 0.01      & 1.9      $\pm$ 0.07    & 0.08             $\pm$ 0.0             & 1.85           $\pm$ 0.01          \\
        $Windmi\_8$             & 0.08      & 2.41      $\pm$ 1.02          & 0.08       $\pm$ 0.01      & 1.99     $\pm$ 0.04    & 0.07             $\pm$ 0.01            & 1.96           $\pm$ 0.03          \\
        $Rucklidge\_9$          & 0.08      & 3.28      $\pm$ 0.61          & 0.2        $\pm$ 0.01      & 1.94     $\pm$ 0.05    & 0.21             $\pm$ 0.01            & 1.94           $\pm$ 0.03          \\
        $Chen\_10$              & 0.005      & 2.61      $\pm$ 0.18          & 1.85       $\pm$ 0.31      & 2.12     $\pm$ 0.09    & 1.94             $\pm$ 0.1             & 2.12           $\pm$ 0.01          \\
        $Windmi\_11$            & 0.08      & 2.43      $\pm$ 0.77          & 0.07       $\pm$ 0.01      & 1.95     $\pm$ 0.05    & 0.07             $\pm$ 0.01            & 1.96           $\pm$ 0.03          \\
        $C. Butterfly\_12$  & 0.08      & 4.86      $\pm$ 0.14          & 0.13       $\pm$ 0.01      & 2.21     $\pm$ 0.04    & 0.17             $\pm$ 0.0             & 2.25           $\pm$ 0.01          \\
        $Halvorsen\_13$         & 0.02      & 1.63      $\pm$ 0.07          & 0.69       $\pm$ 0.05      & 2.11     $\pm$ 0.0     & 0.79             $\pm$ 0.02            & 2.11           $\pm$ 0.0           \\
        $Rucklidge\_14$         & 0.08      & 3.25      $\pm$ 0.62          & 0.19       $\pm$ 0.01      & 1.93     $\pm$ 0.05    & 0.21             $\pm$ 0.01            & 1.94           $\pm$ 0.03          \\
        $Lorenz\_15$            & 0.02      & 4.75      $\pm$ 1.05          & 0.8        $\pm$ 0.03      & 2.06     $\pm$ 0.01    & 0.84             $\pm$ 0.01            & 2.06           $\pm$ 0.01          \\
        \bottomrule
    \end{tabular}
    }
  \caption{\textbf{Performance metrics for the 16 overlapping attractor task.} Measurements performed on 100 predicted trajectories (each using a different random seed for the ERT regressor) and 100 test time series initialized from distinct starting points.}
  \label{tab:16}
\end{table}

\newpage
\subsection*{Failure modes from defects in short-term memory processing}

\begin{figure}[t]
    \centering
    \includegraphics[width=0.83\linewidth]{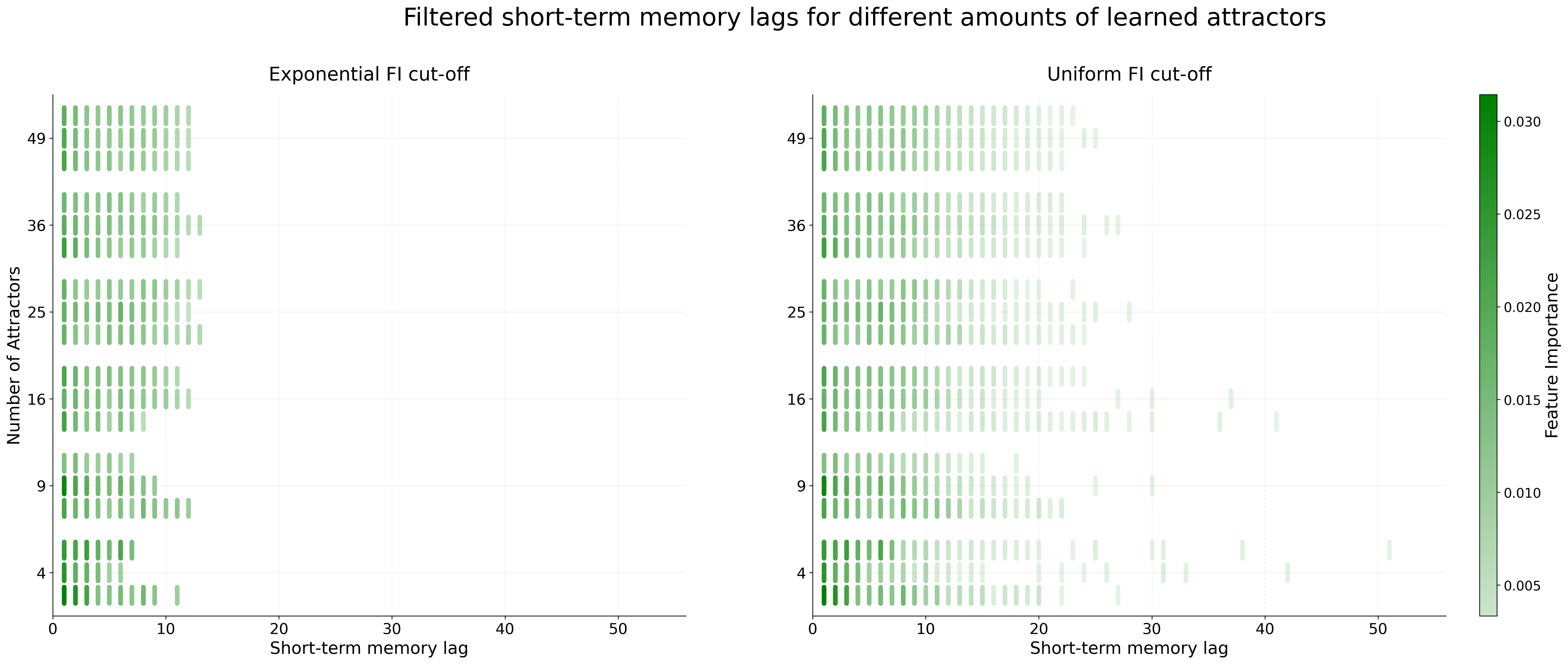}
    \caption{\textbf{Filtered short-term memory representation for different attractor tasks.} Left panels show filtered short-term memory representation using exponential feature importance filtering for varying numbers of learned attractors, displayed for z (top), y (middle), and x (bottom) coordinates. Right panels show corresponding results using uniform filtering. Each lag represents one time step in the past.}
    \label{fig:short-term}
\end{figure}

%{\af\em(improved but still not completely clear to me what's going on here)}
%{\af \st{In this subsection we investigate the retained feature configuration obtained after training NGRC+ERT systems in a VAR setup to reconstruct different numbers of attractors. } 
In this subsection we conduct a more detailed analysis on the retained features, i.e., features that remain after filtering. 
%{\dk \em(and how these influence the performance of the NGRC+ERT system when trained to reconstruct multiple different attractors)}
More specifically, we examine how the retained features are distributed across the time-delay features of the x, y, and z coordinates after applying (i) the exponential filtering scheme and (ii) the uniform filtering scheme. We consider six sub-tasks involving the reconstruction of 4, 9, 16, 25, 36, and 49 overlapping attractors, respectively. The hyperparameter configuration is fixed across all sub-tasks to $k=100$, $s=1$, $\gamma=25$ and $T=20{,}000$ for each attractor. 
%More specifically, we examine {\af\st{in the multiple-overlapping-attractor task}} how the retained features are distributed across the constituent $z$, $y$, and $x$ coordinates and their corresponding time delayed features, i.e., $z(t),~ z(t-1), ~\ldots$, after training with each filtering scheme to reconstruct 4, 9, 16, 25, 36 and 49 overlapping attractors, thus specifying six distinct sub-tasks.We fix the hyperparameter configuration across all sub-tasks to $k=100$, $s=1$, $\gamma=25$ and use a training size of $T=20{,}000$ for each attractor.
In Figure~\ref{fig:short-term} we show the optimized feature configurations for each sub-task, by decomposing the retained features into the constituent z (top), y (middle), and x (bottom) coordinates. We refer to this decomposed representation as the optimized short-term memory processing being performed by the NGRC+ERT system.
%and specify that the short term memory lag of 1 corresponds to the processing of the most recent input data at time $t$ and lag 2 to that at time $t-1$ and so on. 
Additionally, we display for each retained feature the obtained feature importance in green where the darker the shade the greater the importance.
We find that the feature importance measure tends to assign the largest importance to features closest to the current time for both schemes.
%,{with the importance decreasing {\af\st{for more temporally remote memories}for features further away from the current time}}.
%{\af\st{The uniform cut-off}
For the exponential filtering scheme, we find that retained features tend to be similar across each sub-task and do not involve time-delay terms greater than 15, i.e., no features beyond $z(t-15)$ are used. In contrast, for the uniform filtering scheme, we find a more diverse range of retained features across each sub-task with much larger time-delay terms in used, even $z(t-51)$ is deemed to be of importance.
% yields more and, particularly for tasks with fewer attractors, also temporally disconnected features.
%{\dk \st{{\af In terms of the uniform filtering scheme we find}, particularly for tasks with few attractors, that there are active features at various times away from the current time, while the exponential filtering scheme maintains {\af{continuous short-term memory} what does this mean??} and focuses the attention of the system on fewer and the most recent features across all attractor numbers.}}
%{\af\em(Last sentence too difficult to follow, we need to discuss this before the paper can be submitted.)}}

\begin{figure}[t]
    \centering
    \includegraphics[width=\linewidth]{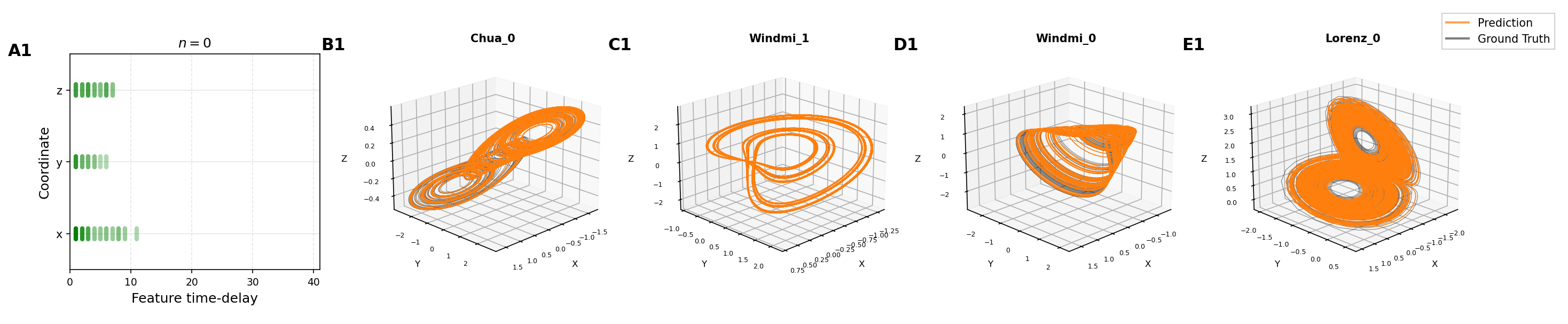}
    
    \vspace{0.5cm}
    
    \includegraphics[width=\linewidth]{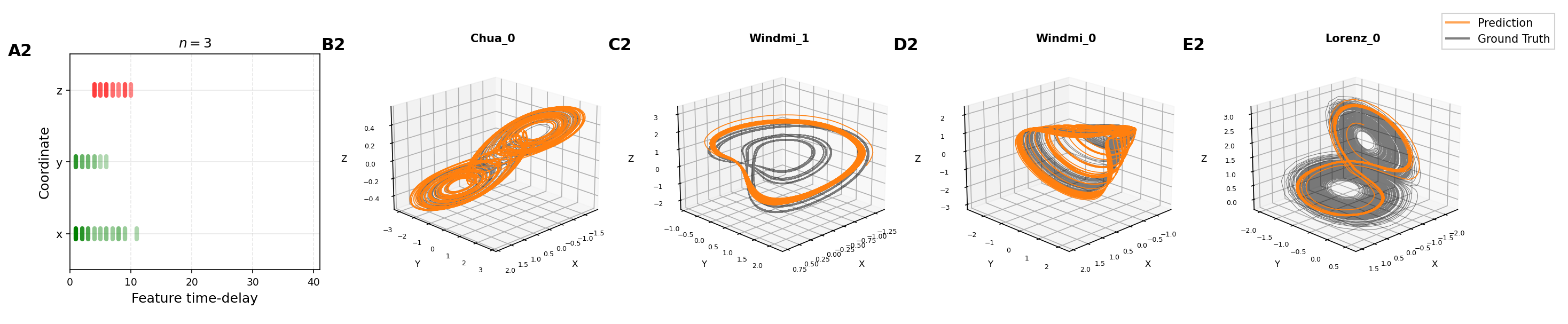}
    
    \vspace{0.5cm}
    
    \includegraphics[width=\linewidth]{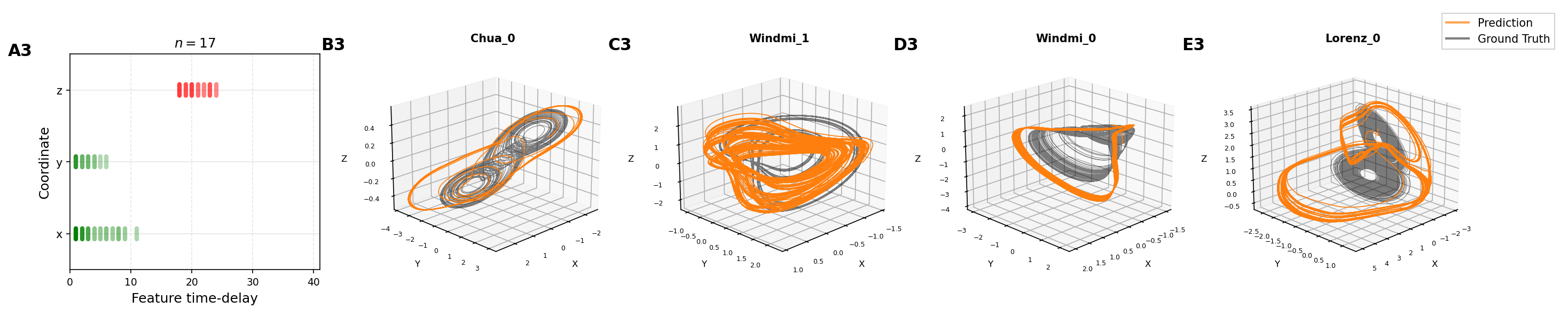}
    \caption{\textbf{Two-shot learning of four attractor system and processing delay effects.} \textbf{Top:} Optimized short-term memory representation obtained from two-shot learning using exponential filtering. Panels show the four-attractor components: chaotic Chua, periodic Windmi, chaotic Windmi, and chaotic Lorenz. \textbf{Middle:} Lag 3 - Processing defects in the z coordinate induce transitions to generated attractors during long-term memory recall. \textbf{Bottom:} Lag 17 - Increased processing defects generate diverse generated attractor behaviors during recall.}
    \label{fig:4attractor}
\end{figure}

\begin{figure}[H]
    \centering
    \includegraphics[width=1\linewidth]{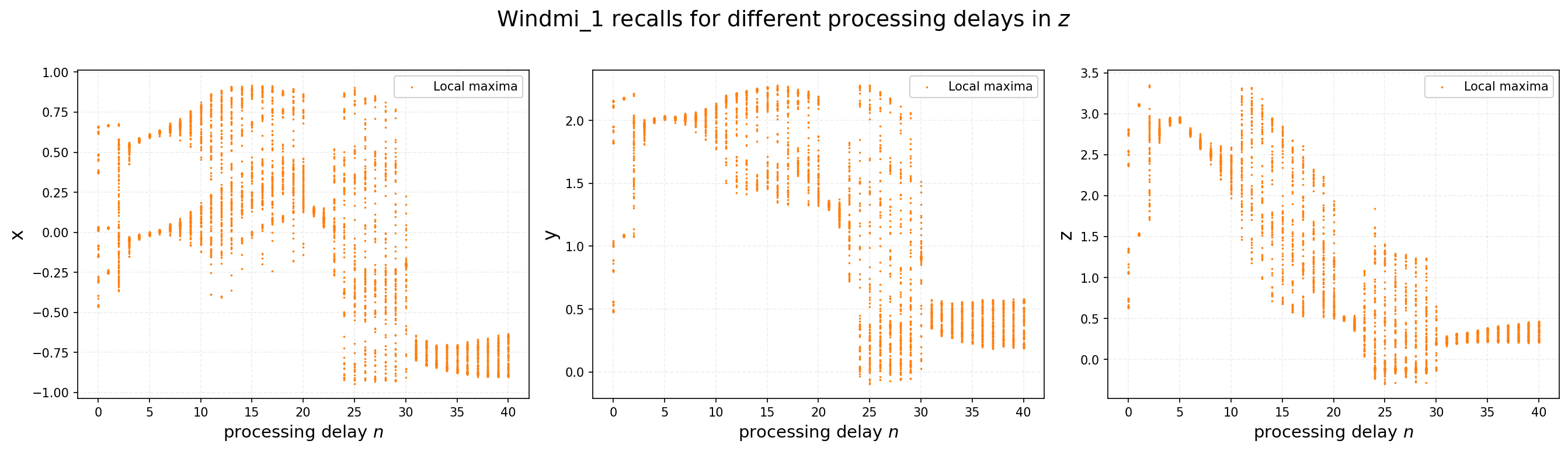}
    \caption{\textbf{Attractor recalls of Windmi\_1 for processing delays.} Long-term memory recall under varying processing delays in $z$ reveals rich generated dynamical behaviors including periodic and chaotic regimes.}
    \label{fig:4attractorBif}
\end{figure}

From here we analyze how defects in short-term memory processing influence long-term memory recall.  By defects, we mean that features are placed at incorrect positions within the retained feature vector the ERT was trained with to perform long-term memory recall. We conduct the analysis by keeping the trained ERT unit fixed, while varying the feature configuration.
This experimental paradigm models how variations in short-term memory processing can induce untrained behavior in learning systems, with implications for hardware-level applications of our approach.
% {\af\st{ signal processing in neuromorphic, or even biological, learning systems}}.
%{\af\em(how do you know they are placed at incorrect positions?)}{\dk \em  e.g. if at the location of feature $f_1$, the values of feature $f_3$ are processed, then the value of $f_3$ are at an incorrect position, hence are wrongly processed, this is what we simulate|| this is a concept, with processing being a subtyp, wrong assignments across differnet coordinates would be different typ}
We investigate the 4-attractor sub-task with the exponential filtering scheme applied (Figure~\ref{fig:4attractor}). Starting from the retained feature configuration (Figure~\ref{fig:4attractor} \textbf{A1}), we simulate processing delays of $n$ time steps for features associated with $z$, e.g., feature $f_{j}$ processes $z(t-n)$ instead of $z(t)$.
% , $f_{j-3}$ processes $z(t-1-n)$ instead of $z(t-1)$ and so on.
%{\dk Specifically, we shift the retained features of the $z$-coordinate by $n$ time-lags, so that the system processes the features of $z(t-n+1)$ instead of $z(t)$. The same processing delay is applied to all retained features associated with $z$, as illustrated in red in the left plots of Figure~\ref{fig:4attractor}.}
%Specifically, we shift the memory of the $z$-{\dk \st{coordinate}variable} by $n$ lags, such that the system processes the features of $z(t-n+1)$ instead of $z(t)$, with corresponding delays applied to all retained features in $z$ as illustrated in red within the left plots of Figure~\ref{fig:4attractor}. 
%Crucially, the ERT was trained only on {\af{synchronized input coordinates}\em(what does this mean?)}. 
We find that by introducing this defect, the system starts recalling deformations of the attractors it was trained on, exhibiting confabulations in the form of `generated attractors'\cite{flynn2025confabulation}.
%{\af\em(but so what? the system was also not trained to reconstruct any attractors with these features?) \dk 1) This experimental paradigm models temporal processing defects in learning systems, where different input channels may exhibit variable transmission delays. 2) Our results reveal that temporal delays in signal transmission can induce rich bifurcation behavior, in which the initially optimized functioning of the learning system breaks and initiates confabulations 3) Is possible in biological neurons, where methlyinzation speeds up neural pathways, or vise versa leads to slower pathways that initiates processing delays within signal processing, hence long-term memory recall is affected, like we show here)}
Figure~\ref{fig:4attractor} (middle and bottom row) shows the predicted dynamics with processing delays of $n=3$ and $n=17$. 
Comparing some of the changes from the top row, in Fig.~\ref{fig:4attractor} \textbf{C2} we see the inner loop of the Windmi\_1 attractor vanishes and the period of the limit cycle reduces, in contrast, in Fig.~\ref{fig:4attractor} \textbf{C3}, the Windmi$\_$1 attractor displays chaotic behavior. Key changes to the other attractors also occur for changing $n$.
For the Windmi$\_$1 attractor we systematically analyze the dynamics for increasing processing delays up to $n=40$ and visualize in Fig.~\ref{fig:4attractorBif} the attractor recall by plotting the changes in the local maxima versus processing delay $n$, see Fig.~S11 in the Supplementary Information for corresponding state space diagrams.
Interestingly, near $n=3,~ 20,$ and $30$, we observe a major change in the generated dynamics; specifically, a sudden contraction/expansion in the size of the attractor, which is indicative of an interior crisis taking place.
%{\af{attractor recalls}\em(predictions?) by plotting the changes in the local maxima and minima versus time lag \st{using a bifurcation diagram, with the time lag being the bifurcation parameter} \em(these are not bifurcation diagrams per se, we can see bifurcations taking place but the actual bifurcations which underlie the changes seen here are not identified)}.
We find that these defects can result in various 
%{\af{processing delays}\em(what is a processing delay?)} {\af\st{can force rich sets of} can result in various}
generated attractors, including periodic and chaotic attractors, that the system was never trained on. 
Further, interior crises commonly take place, thus highlighting the role played by unstable equilibria in state space whose presence we observe indirectly.  %{\af\em(sentence not clear)}.
In Supplementary Figures S11-S12, we observe similar behavior when processing delays are applied to two retained features across all coordinates.
%In Supplementary Figures S10-S11 we observe similar behavior when {\dk \st{memory} retained} features associated across all coordinates features  experience {\dk processing delays} {\af\em(what is a memory feature?)}.
%{\dk Strong statements: }Such processing defects could potentially represent various biological phenomena, like synaptic transmission delays caused by amyloid plaques in Alzheimer's disease, differential processing speeds between frequently used and underutilized neural pathways, or variations in signal propagation due to differences in axonal myelination.{\dk <- needs citations..lots }

%{\dk \st{This experimental paradigm models temporal processing defects in learning systems, where different input channels may exhibit variable transmission delays {\af\em(sentence not clear whatsoever)}.} Our results reveal that processing delays in {\af{signal transmission} \em(where is this defined?)} can induce rich bifurcation behavior, in which the {\af{initially optimized functioning of the learning system breaks}\em(no idea what this means)} and initiates confabulations\cite{flynn2025confabulation}, highlighting the importance of accurate {\af{temporal memory processing} \em(???)} within learning systems.  -> in discussion }

\newpage
\section*{Discussion}

% {\af 
In this paper we extend a two-shot learning approach introduced by Giammarese \textit{et al.}~\cite{giammarese2024tree} to train a single machine learning system to store and recall multiple strange attractors.
In short, we apply this approach to a system consisting of a short-term and long-term memory unit, specifically a next generation reservoir computer (NGRC) and an extremely randomized tree (ERT), and take inspiration from multifunctional and parameter-aware reservoir computing\cite{flynn2021multifunctionality,flynn2022exploring,flynn2023seeing,flynn2024switching,kim2021teaching,kong2021machine,kong2023reservoir,koglmayr2024extrapolating} to train the combined NGRC+ERT system to reconstruct multiple strange attractors. The two-shot learning approach significantly improves performance without the need for extensive hyperparameter optimization by (i) identifying important features of the NGRC+ERT system and (ii) training a new NGRC+ERT system based on these features.
The separation between short-term memory processing (NGRC) and long-term memory recall (ERT) also enables greater interpretation of the two-shot learning. We show how the information obtained from the first-shot informs the second, specifically, after an initial optimization we find subsequent training focuses exclusively on short-term memory features that prove important for the given task, thereby improving performance without further hyperparameter tuning.

Our results shed light on the relationship between short- and long-term memory and how they work together to store and recall memories, in our case several different strange attractors. More specifically, we study the factors that disrupt this relationship by highlighting that stable storage and recall of multiple attractors requires clear separation in feature space. When the separation is insufficient, the NGRC+ERT system exhibits a tendency to recall or `switch between' different trained attractors. We show that this behavior can be prevented through adequate feature separation. Building on this, we investigate recall properties when defects are introduced during short-term memory processing. We find that the system exhibits momentary confabulations in the form of generating attractors, attractors that partially resemble the original attractor. This is a qualitatively different outcome than recalling the wrong trained attractor, rather the NGRC+ERT confabulates in the sense of O'Hagan \textit{et al.}~\cite{flynn2025confabulation}.

% {\af While our experiments are based on artificial data, chaotic attractors generated by noise-free dynamical systems, our results serve as a proof of concept and may guide future work based on real-world or noisy data.}
% {\af Moreover, } 
Our findings may inspire future studies on the relationship between short- and long-term memory in other learning systems, like classic reservoir computers, where short-term memory processing is facilitated through a network structure and influenced by node saturation through the activation function. % {\af(\em I haven't written the paper on this yet ...)}. 
Our approach performs long-term memory recall through trees, whereas classical reservoir computing compresses this task into a matrix typically obtained from ridge regression. These represent qualitatively different learning methodologies as trees can grow dynamically, while matrices are finite-dimensional. However, it is important to note that despite their structural differences, these methodologies exhibit similar failure modes, like the emergence of generated attractors and switching dynamics. This may point towards the existence of a more universal principle governing failure modes in learning systems. 

\section*{Methods}

%{\af rough work: NGRC is a fast-paced and exciting development in ML. It offers X, Y, Z over classic RC methods, and is deeply rooted other ML areas, eg. VAR. Recently Bollt showed that instead of training an NGRC using the same techniques as classic RCs, i.e., ridge regression, one can train the NGRC using 'tree based methods' (combining another well-developed ML area). Bollts results showed X, Y, Z in comparison to ridge regression. However, there are certain shortcomings to Bollts results that we address in this paper. Our study marks several key advancements from Bollts work: (i) the extension from 1-d data, (ii) the extension to multifunctionality, and (iii) provides clarity on why Bollts found VAR features to provide the best performance by comparing it with other feature selection methods. from a conceptual standpoint, we also develop the notion of short and long-term memory separation between the different component that we use to construct our NGRC+ERT system, specifically, we relate the concept of different brain regions that are responsible for different memory tasks (short and long-term) to the properties of our system.}

\subsection*{Extremely randomized trees}

%{\af \em (A companion figure which explains the process below would be helpful. Consider a figure consisting of subfigures that show in (a) the process of constructing trees, (b) the process of constructing ERTs which also shows the processes in  Eq. 1 and 2.)}

% This subsection introduces the main concepts behind the Extremely Randomized Trees (ERT) algorithm~\cite{geurts2006extremely}, which is a type of decision tree machine learning-based approach for classification and regression tasks.

This subsection introduces the key principles of the Extremely Randomized Trees (ERT) algorithm~\cite{geurts2006extremely}, a machine learning-based approach for solving classification and regression tasks.

% {\af In general, these tasks consist of constructing `features' from training data, i.e., manipulating training data according to a specified method, and finding a function which maps these features to a target.}

\subsubsection*{Overview of decision trees}

A \textbf{decision tree} is a non-parametric model made up of hierarchically structured \textit{nodes} connected by \textit{branches} that represent if-then-else rules. We introduce the following notation to describe a decision tree in greater detail. %{\af We need to introduce the following notation to describe a decision tree in greater detail.}
Let $D = \{(\mathbf{r}_i, \mathbf{y}_i) \}_{i=1}^{T}$ be the training data set, with $T$ elements. Elements of this set, known as training samples, consist of the following pair, a feature vector, $\mathbf{r}_i \in \mathbb{R}^Z$, where $Z \in \mathbb{N}$ is the number of features, and a target vector $\mathbf{y}_i \in \mathbb{R}^d$, where $d \in \mathbb{N}$ is the number of components of the vector. %vectors do not have a dimension
A feature is a quantity derived from elements of the training data set and used as input to, in this instance, the decision tree. 
%In this work, features are constructed using the NGRC framework described later in the Methods section. 
To simplify the notation below, we define a set of indices $\mathcal{I} = \{1,\dots,T\}$ to reference the training samples, where $i \in \mathcal{I}$ references the training sample $(\mathbf{r}_i, \mathbf{y}_i)$. A tree starts with a single node, called the \textit{root node}, which represents the entire set of indices $\mathcal{I}$, hence the entire training data set.
\\
\textbf{Training the tree: } Starting from the root node, the tree `grows' by recursively partitioning the training data set into two smaller subsets, forming the left and right child nodes. Each child node is split again, and the process continues until a stopping criterion is reached, such as a maximum depth of the tree, defined as the number of instances that new branches are created, or a minimum number of training samples per node. A node that is not split further is called a leaf node, or simply a leaf. Each leaf stores a subset of training samples, identified by their indices. Once the stopping criterion is reached, the tree-building process is complete, i.e., the tree is trained. The splitting rule at each node depends on the specific tree method, we introduce the splitting rule for ERT later.
\\
\textbf{Making predictions with the tree: } Given a new input $\mathbf{r} \in \mathbb{R}^Z$, the tree makes a prediction by `routing' $\mathbf{r}$ from the root node through the tree's branches to the respective leaf. More specifically, at each node, the splitting condition learned during the tree building process determines whether $\mathbf{r}$ is forwarded to the left or right child node. This process continues until a leaf node is reached. 
% {\dk \st{This process is captured in the mapping $tree(\text{input)} = \text{prediction}$.}}
In a regression setting, the prediction of a tree for an input $\mathbf{r}$ is the arithmetic mean of the target vectors of the training samples that reached the same leaf. If $\mathbf{r}$ reaches leaf $l$ the prediction is given by,
\begin{equation} \label{eq:single_tree_prediction}
    tree(\mathbf{r}) = \frac{1}{|\mathcal{I}_l|} \sum_{i \in \mathcal{I}_l} \mathbf{y}_i = \mathbb{E}[\mathbf{y_i}\mid i \in \mathcal{I}_l] = \mathbf{\hat{y}},
\end{equation}
where $\mathcal{I}_l$ denotes the set of indices of training samples assigned to leaf $l$. $\mathbb{E}$ denotes the arithmetic mean of the target vector samples in $\mathcal{I}_l$. Specifically, the sum is element-wise for target vectors with $d$ components.

\begin{figure}[t]
  \centering
  \includegraphics[width=0.85\textwidth]{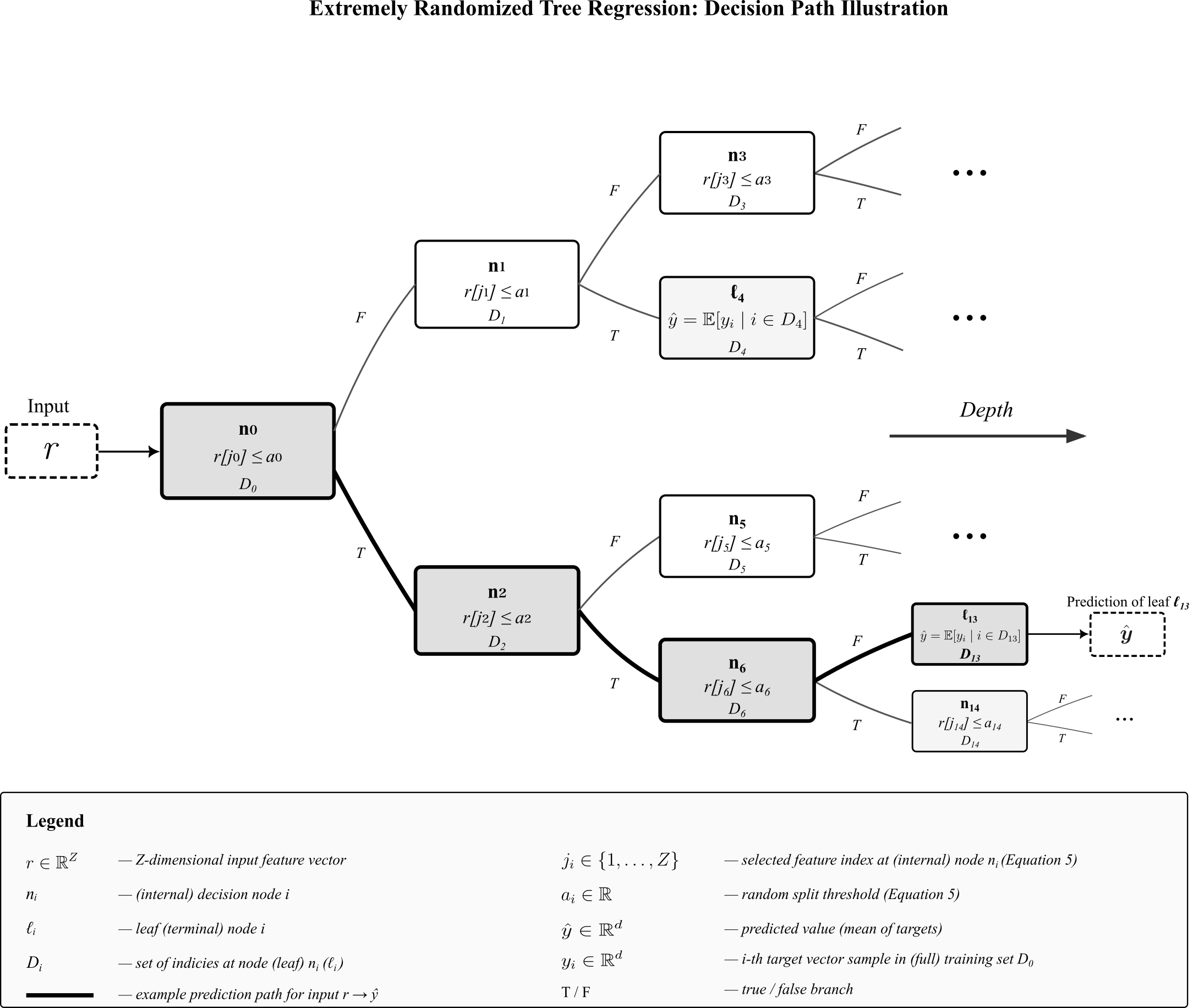}
  \caption{\textbf{Structure of an extremely randomized tree.} The tree receives an input $\mathbf{r}$, which traverses the branches by comparing input values with feature thresholds at each node until reaching a leaf, where it outputs a prediction $\mathbf{\hat{y}}$.}
  \label{fig:extra-trees-pred}
\end{figure}
%However, the issue with many tree based methods is high variance in their prediction\cite{geurts2006extremely}, which is a consequence of the hierarchical structure. Trees generate splits which depend on the training data samples, a slight perturbation in the training samples can lead to different splits and this difference propagates further down the tree. This leads to different predictions, thus making a single tree less robust with respect to fluctuations in the data.
%{\af\em(At present, the motivation to use ERT is that `classic trees' are not robust to fluctuations in the training data samples. Can you please clarify what fluctuations are present in the training data samples that are used in the experiments and state why `classic trees' are not suitable to use in this instance.)(should the motivation also be somehow that `classic trees' are unable to be used in the multi-attractor reconstruction style tasks explored in the paper?)(should there also be motivation that `classic trees' do not have the same long-term memory capabilities as ERTs?)(it would be good to discuss and clarify the above.)}
%\textbf{Description of `extremely randomized trees' in machine learning: } 
%{\af\em(many concepts in this section are still not clear to me)}
\subsubsection*{Overview of extremely randomized trees}
\textbf{Extremely randomized trees (ERT): } Many tree-based machine learning methods suffer from high prediction variance~\cite{geurts2006extremely}, a consequence of their hierarchical structure. As the splitting process depends on the training data, small perturbations in the training samples can produce different splits, and these differences propagate through the tree, making the learning process sensitive to small changes in the training data. 
%{\ms In our case, the training data consists of chaotic time series, which are known to be highly sensitive to initial conditions, leading to different trajectories. \emph{Empirically, the prediction of single decision trees is susceptible to these changes, leading potentially to worse results under the same hyperparameter settings. (Do not have a statistics on that, might be requested in a review? Maybe leave out then?)}}
The ERT algorithm trains an ensemble of independent trees, averaging their predictions after training to address this issue. 
\\
%\textbf{Training extremely randomized trees: } The ERT algorithm builds $M$ trees independently, where each tree is built using the same set of rules, but introduces randomness at each split so that the final trees differ.{\af{\em(this sentence does not make sense, how is D given? avoid `by some means' as one can confuse this means with the arithmetic mean ... better to avoid using this phrase here )}{Following the notation above, a set of sample indices $D$ is given, where the indices refer to the training samples in the training data set, which consist of features constructed by some means, in our case by the NGRC framework, and target vectors.}}
\textbf{Training extremely randomized trees:} %{\dk quite some changes here in the mathematics, I commented out older parts }
The ERT algorithm builds $M$ trees independently. Each tree follows the same construction rules but introduces randomness at each split, resulting in $M$ distinct trees.
Let $D = \{(\mathbf{r}_i, \mathbf{y}_i)\}_{i=1}^{T}$ denote the training data set, where $\mathbf{r}_i \in \mathbb{R}^{Z}$ is the feature vector and $\mathbf{y}_i \in \mathbb{R}^{d}$ is the target vector for the $i$-th sample. Equivalently, we define the feature matrix $\mathbf{R} \in \mathbb{R}^{Z \times T}$ and the target matrix $\mathbf{Y} \in \mathbb{R}^{d \times T}$, where $Z$ is the number of features, $T$ is the number of samples, and $d$ is the dimension of the `target vector space', with $Z, T, d \in \mathbb{N}$. In our case, the feature vectors are constructed using NGRC.
During tree construction, the training data set is partitioned into subsets, each tracked by a corresponding index set. Let $\mathcal{I} \subseteq \{1, \ldots, T\}$ denote such an index set. At each node $n$, let $T' = |\mathcal{I}|$ denote the number of training samples that reach this node, and let $\mathbf{R}' \in \mathbb{R}^{Z \times T'}$ denote the submatrix of $\mathbf{R}$ restricted to these samples. The splitting rule at each node is defined as follows.
From the $Z$ features, $K \leq Z$ features are randomly selected, indexed by $\{f_1, \ldots, f_K\}$. For each selected feature $f_i$, let $\mathbf{R}'_{f_i} \in \mathbb{R}^{1 \times T'}$ denote the corresponding feature row of $\mathbf{R}'$. A random threshold $a_i$ is then drawn uniformly from the interval $[\min(\mathbf{R}'_{f_i}), \max(\mathbf{R}'_{f_i})]$, yielding $K$ candidate splits $\{(f_1, a_1), \ldots, (f_K, a_K)\}$. Each candidate split partitions the training data at node $n$ into two subsets:
\begin{equation} \label{eq:child_nodes_datasets}
    D^{(L)}_i = \{(\mathbf{r}_j, \mathbf{y}_j) \mid j \in \mathcal{I}, \, \mathbf{R}_{f_i,j} \leq a_i\}, \qquad D^{(R)}_i = \{(\mathbf{r}_j, \mathbf{y}_j) \mid j \in \mathcal{I}, \, \mathbf{R}_{f_i,j} > a_i\},
\end{equation}
corresponding to the left (L) and right (R) candidate child nodes. Let $T^{(L)}_i = |D^{(L)}_i|$ and $T^{(R)}_i = |D^{(R)}_i|$ denote the number of samples in each candidate subset, with $T^{(L)}_i + T^{(R)}_i = T'$.
%\st{At each node, a random subset $S$, which consists of $K$ features, is selected. For each feature, $f_i \in S$, where $i\in{1,\dots,K}$, a threshold value $a_i$ is drawn from a uniformly random distribution of values between the minimum and maximum value of that feature within the node} {\af within the random subset $S$?}. \st{Each candidate split $(a_i, f_i)$} {\af\em(it is not clear what this notation means)} \st{divides the data samples into two subsets} {\af of $S$ or of the entire data set? what data samples?} \st{, $D_1$ and $D_2$, and the resulting split is evaluated using a loss function. In our case, the loss is measured using the mean squared loss}:
Each candidate split is evaluated using the squared error loss,
\begin{equation} \label{eq:tree_loss}
    L(f_i, a_i) = \sum_{j \in \mathcal{I}^{(L)}}{(\mathbf{y}_j-\mathbf{\hat{y}}_L)^{2}} + \sum_{j \in \mathcal{I}^{(R)}}{(\mathbf{y}_j-\mathbf{\hat{y}}_R)^{2}},\,
\end{equation}
where $\mathbf{\hat{y}}_L,\, \mathbf{\hat{y}}_R \in \mathbb{R}^d$ are the mean target vectors in each subset defined by
\begin{equation}
    \mathbf{\hat{y}}_p = \frac{1}{T_p} \sum_{j \in \mathcal{I}^{(p)}} \mathbf{y}_j\,,\quad p = L,\,R.
\end{equation}
Minimizing this loss is equivalent to maximizing the variance reduction.
The best splitting pair and with that the learned splitting rule at node $n$ is
\begin{equation} \label{eq:ERT_splitting_rule}
    (f^*, a^*) = \arg\min_{i \in \{1,\dots,K\}} L(f_i, a_i).
\end{equation}
%{\af\em(are these $x_i$ the same as the feature vector and $y_i$ the same target vector as introduced in the description of trees section?)}.
%The optimal splitting rule for a node $n$ is expressed in equation \ref{eq:ERT_splitting_rule}.
%{\af\em(it is not clear what this notation means, where is $a_{split}$ defined?)}. 
%The \textit{left} and \textit{right} child node are created in this manner, representing the set of indices from equation \ref{eq:child_nodes_datasets}, given the optimal feature index.
The partitioning is recursively repeated at each new node until the tree's stopping criterion is reached.
\\
\textbf{Making predictions with the extremely randomized tree: }
Once all $M$ trees are constructed, they can be used for prediction. For an input $\mathbf{r} \in \mathbb{R}^Z$ passing through node $n$, let $(f_{j_n}, a_n)$ denote the splitting pair selected during training, where $j_n \in \{1,\dots,Z\}$ is the index of the selected feature. If $\mathbf{r}[j_n] \leq a_n$, the input is routed to the left child, otherwise, it is routed to the right child. This process continues recursively until the input reaches a leaf node, where the prediction is given by  Eq.~\ref{eq:single_tree_prediction}. Each of the $M$ trees independently process the input $\mathbf{r}$ and produces a prediction. The final ERT prediction is the average across all trees:
\begin{equation}
    \text{ERT}(\mathbf{r}) = \frac{1}{M} \sum_{m=1}^{M} tree_m(\mathbf{r})
\end{equation}
A schematic of the above process for one extremely randomized tree is provided in Fig.~\ref{fig:extra-trees-pred}.
\\
\textbf{Implementation: }
We implement the ERT algorithm using the \texttt{scikit-learn} python library~\cite{scikit-learn}.
%The algorithm has several key hyperparameters; the number of trees $M$, the number of randomly selected features $K$ considered at each split, and the tree depth, which can be controlled either by setting a maximum depth or by specifying the minimum number of samples required at each node.
%In our experiments, these parameters are fixed rather than optimized {\af\em(one arouses less suspicion if the following phrase is used, 'It was found empirically that, ...')}.
We use default tree hyperparameters, except for setting the number of trees to $M = 100$ for two-attractor tasks and $M = 50$ for multiple-attractor tasks, and $K = Z$. The tree depth is not fixed, which leads to fully grown trees within the ensemble.

\subsection*{Feature Importance}
\begin{figure}[t]
  \centering
  \includegraphics[width=0.85\linewidth]{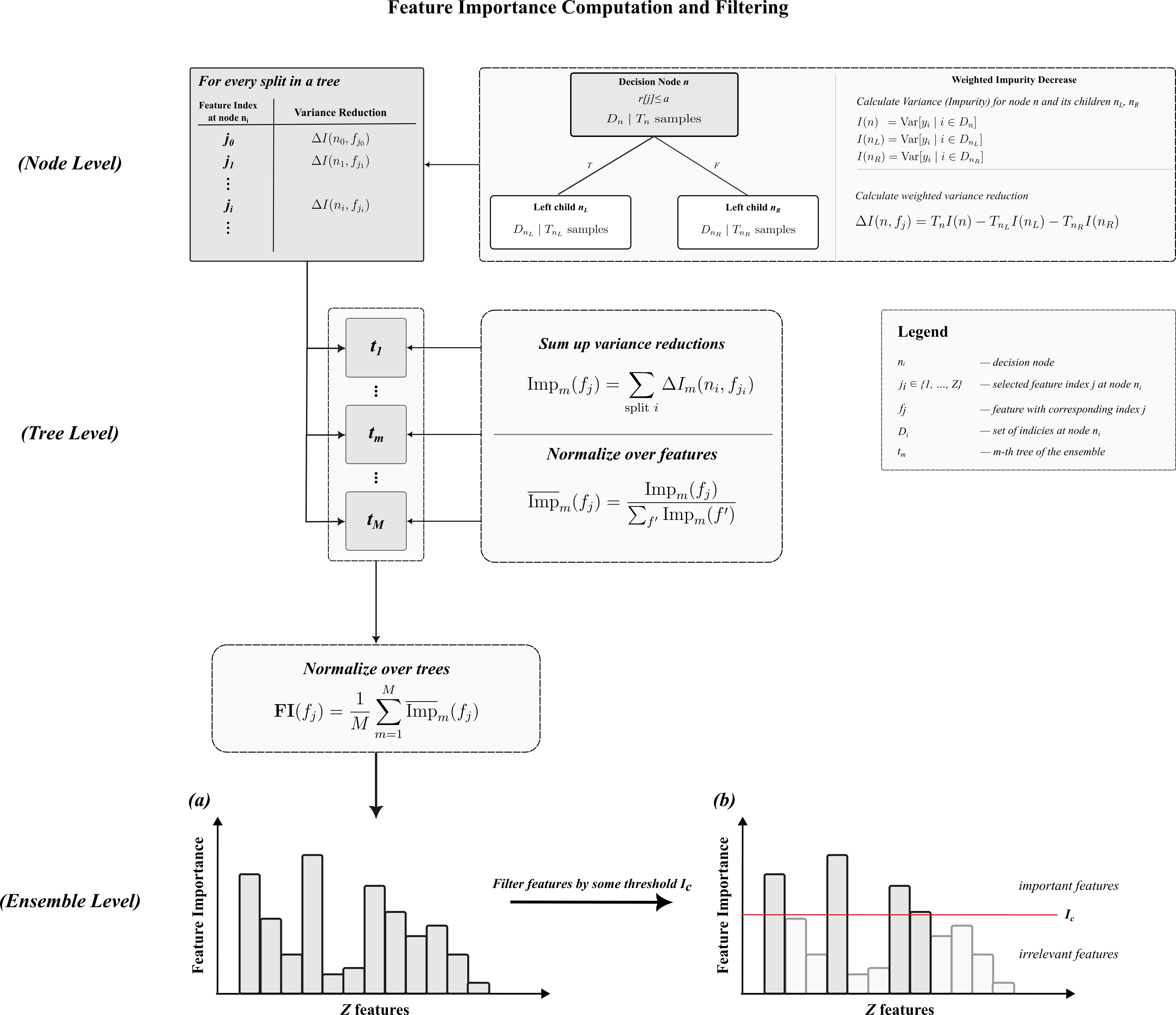}
  \caption{\textbf{Feature importance computation and filtering in Extremely Randomized Trees.} At each node, the weighted variance reduction is calculated for every split. These values are summed across the tree and normalized so that the total importance equals one. At the ensemble level, importance measures are averaged across all trees, yielding a final feature importance value for each feature. Features are then classified as important or irrelevant by comparison with a threshold $\mathrm{I}_c$.}
  \label{fig:feature_importance_NEW}
\end{figure}

%{\af \em (A companion figure which explains the process below would be helpful.)} The `importance of a feature'~\cite{breiman2001random,geurts2006extremely,altmann2010permutation,scikit-learn} quantifies how much that feature contributes to reducing the loss function in  Eq.~\ref{eq:tree_loss} across all splits in the ensemble. We use this measure to optimize the feature configuration provided by the NGRC framework, thereby eliminating the need for extensive hyperparameter tuning typically performed to o feature {\af\em(Insert sentence here which describes why feature importance is computed, i.e., used in the two-shot training of the NGRC+ERT system and the two-shot system is much better than one-shot etc. ... )} {\af To describe the process used to compute feature importance, we first} recall that at each node $n$,
%{\dk commented out older parts}
The \textit{feature importance}~\cite{breiman2001random,geurts2006extremely,altmann2010permutation,scikit-learn} quantifies how much each feature contributes to reducing the loss function in Eq.~\ref{eq:tree_loss} across all splits in the ensemble. We use this measure to optimize the feature configuration of the NGRC framework, eliminating the need for extensive hyperparameter tuning typically required to obtain an optimized feature configuration.
%Specifically, we employ a two-shot learning methodology: first, we train an initial ERT instance and compute the feature importances across the ensemble. Then, we filter only those features whose importance exceeds a specified threshold and use the filtered feature configuration to train a second ERT instance.
To describe how feature importance is calculated, we first recall that at each node $n$, the selected split $(f_j, a_j)$ partitions the samples into child nodes $n_L$ and $n_R$. 
%This variance reduction forms the basis for measuring feature importance. Consider a non-leaf node $n$ that is split on feature $f_j$ with threshold $a_j$, producing the left and right child nodes $n_L$ and $n_R$. 
Let $T_n$, $T_{n_L}$ and $T_{n_R}$ denote the number of samples in these nodes and let $I(\cdot)$ denote the \textit{impurity at a given node}, which is the variance of the target values within these nodes.
The contribution of the split to the feature importance calculation of $f_j$ is the weighted variance reduction,
\begin{equation}
    \Delta I_m(n, f_j) = T_n I(n) - T_{n_L} I(n_L) - T_{n_R} I(n_R).
\end{equation}
This quantity is computed for every split in the $m$-th tree and normalized across all features, such that
%such that the importance is comparable across trees,
\begin{align}
    \text{Imp}_m(f_j) &= \sum_{\text{split } i} \Delta I_m(n_i,f_{j_i}) \label{eq:Imp per tree},\\
    \overline{\text{Imp}}_m(f_j) &= \frac{\text{Imp}_m(f_j)}{\sum_{f'} \text{Imp}_m(f')} \label{eq:normalized imp tree level}
\end{align}
where the sum in Eq.~\ref{eq:Imp per tree} is computed over all nodes split by $f_j$ and Eq.~\ref{eq:normalized imp tree level} specifies the normalization across all features. 
%To obtain the `{\af\em{final}} importance' of feature $f_j$ {\af\em(just call this the feature importance of feature $f_j$?)}, we sum all normalized importances computed in   Eq.~\ref{eq:normalized imp tree level}  and {\af\em{normalize across the ensemble} (do you mean you divide by M? i.e. take the average)}, which yields the feature importance (FI)
To obtain the feature importance of $f_j$, we take the average of the normalized importances computed in Eq.~\ref{eq:normalized imp tree level} across the ensemble, yielding the feature importance (FI)
\begin{equation}
    \textbf{FI}(f_j) = \frac{1}{M}\sum_{m=1}^M \overline{\text{Imp}}_m(f_j).
\end{equation}
A schematic of the above process is provided in Fig.~\ref{fig:feature_importance_NEW}.

\subsection*{Next-Generation Reservoir Computing}

Next-generation reservoir computing is a machine learning approach introduced by Gauthier \textit{et al.}~\cite{gauthier2021next} with applications in, for instance, time series prediction, attractor reconstruction, and control \cite{gauthier2021next,barbosa2022learning,haluszczynski2023controlling}.
%{\af\st{In this paper, we view next-generation reservoir computing as a framework that employs a deterministic structure derived from nonlinear vector autoregression (NVAR) to construct feature vectors from input data.}}
Unlike traditional reservoir computers (RCs) that rely on random networks \cite{jaeger2004harnessing,maass2002real,verstraeten2007experimental}, feature vectors for next-generation reservoir computers (NGRCs) are constructed in a deterministic manner by computing unique monomials of time-shifted input variables. 

While RCs and NGRCs are typically trained via ridge regression, 
% NGRCs have proven effective for time series prediction and attractor reconstruction \cite{gauthier2021next,barbosa2022learning,haluszczynski2023controlling}. 
Giammarese \textit{et al.}~\cite{giammarese2024tree} recently introduced a new technique for training NGRCs which improves their attractor reconstruction capabilities by incorporating an `Extremely Randomized Tree' (ERT) unit as a trainable output layer using their TreeDOX framework (\textbf{Tree}-based \textbf{D}elay \textbf{O}verembedded e\textbf{X}plicit memory learning of chaos). In this paper, we extend the results of Giammarese \textit{et al.}~\cite{giammarese2024tree} by introducing a new technique for training NGRCs to reconstruct multiple attractors from different dynamical systems, thereby extending the TreeDOX framework to multifunctional\cite{flynn2021multifunctionality,flynn2022exploring,flynn2024switching} and parameter-aware (or index-based) learning \cite{kim2021teaching,kong2021machine,kong2023reservoir,koglmayr2024extrapolating}

% {\af\st{NGRC framework to reconstruct multiple attractors from different dynamical systems by incorporating an Extremely Randomized Trees (ERT) unit as a trainable output layer, following the TreeDOX framework \cite{giammarese2024tree}, thereby extending it to multifunctional\cite{flynn2021multifunctionality,flynn2022exploring,flynn2024switching} and parameter-aware (or index-based) learning \cite{kim2021teaching,kong2021machine,kong2023reservoir,koglmayr2024extrapolating}.}}

% For attractor reconstruction tasks, 

\textbf{Training an NGRC system to perform attractor reconstruction: }
% {\af 
The aim in attractor reconstruction tasks is to use a time series describing a trajectory on a given attractor to train some system to reconstruct the future `climate' of the attractor, i.e., mimic the long-term dynamics, while at the same time provide reasonable short-term predictions on the trajectory's future.
% } 
% the feature vector construction process for an NGRC is as follows. 
% {\af 
When using an NGRC for attractor reconstruction tasks, the first step is to construct features using the procedure outlined below.
% }
For a given data point in the `input time series', $\mathbf{X} \in \mathbb{R}^{d \times T}$, containing $T>0$ data points each consisting of $d>0$ components, 
% {\af
which corresponds to discrete-time samples of a trajectory on a given attractor,
% }
we first construct an initial `linear time-delay feature vector' using $k\geq1$ past data points each separated by $s\geq1$ time steps.
%{\af\st{, are concatenated to form a linear time-delay feature vector.}}
%{\af\st{The time-delay embedding requires a} 
Therefore, there is a fixed `warm-up phase' of $\delta t = (k-1) \cdot s + 1$ time steps needed to create the initial feature vector, reducing the effective training data set to $T - \delta t$ samples. 
%{\af\em(should there be some restriction on $k$ and $s$ so $\delta t$ is $\geq 0$ or $\geq 1$? Is it $k,s\geq1$ so $\delta t \geq 1$) \dk might be a physics view, no negative numbers of time steps k possible and no negative spaces s between them as well, so per definition of the parameters explained}
%{\af\em(please specify, the range of values that a given quantity can take should always be specified.) (please go back and check if all integers/real numbers have been defined in this way.)}
The `final feature vector' $\mathbf{r}_i$ at time $t_i$ is obtained by computing unique monomials up to order $O$ from the components of this linear time-delay vector, introducing nonlinearity into the feature construction process. Here, unique means that each monomial term appears exactly once. Thus, {when given $\mathbf{X}$,}
% for a $d$-dimensional time series $\mathbf{X} \in \mathbb{R}^{d \times T}$ corresponding to a trajectory on a given attractor over a duration of time $T$, 
we define the 
% {\af\st{matrix of feature vectors available for training}
`feature matrix'
% }
as,
% We aim to learn and predict the dynamics, thereby evaluating the prediction performance through short-term prediction accuracy (Eq.~\ref{mean_forecast_horizon}) and long-term behavior reconstruction metrics (Eq.~\ref{eq:Long_term_measurement_error}).
\begin{equation}
\mathbf{R}(\mathbf{X}, k, s, O) = [\mathbf{r}_{\delta t}, \mathbf{r}_{\delta t + 1}, \ldots, \mathbf{r}_{T-1}] \in \mathbb{R}^{Z \times (T - 1 - \delta t)},
\end{equation}
where $Z$ denotes the number of features as before. 
% {\af 
Note, there are $T - 1 - \delta t$ columns in this matrix, we explain why below.
% }
%{\af\em(changed as Z was previously used to denote number of features, not feature vector dimension. while the may mean similar things it is best not to redefine terms of use them interchangeably)}
% {\af\st{feature vector dimension}}.
% {\af
The type of features used depend on the choice of $O$ and $k$.
% }
% {\af 
In this paper
% }
we construct features in three distinct ways and refer to each of the resulting sets of features as a `feature model type' to further stress that the features are a `model', i.e., mapping, of the input data; see Figure S1 in the Supplementary Information for an illustration of each feature model type.
% The framework provides three distinct `feature model types' (illustrated in Supplementary Figure S1) {\af and by model we mean mapping of the input data?}.
% {\af 
In short, 
% }
the feature model types are created by (i) setting $O = 1$ and $k>1$, which reduces the NGRC to a vector autoregression (VAR) model containing only linear time-delayed features, (ii) setting $O > 1$ and $k>1$, thereby including higher-order monomials which yields a nonlinear vector autoregression (NVAR) model, and (iii) setting $O > 1$ and $k=1$, which produces a nonlinear memoryless model (NML) that discards temporal information. 
%Given the feature matrix $\mathbf{R}$, the training target matrix $\mathbf{Y}$ specifies the output that the { \st{model} \af\em(what model?)} framework is trained on to predict. We employ the NGRC framework as a one-step-ahead integrator in the following sense, where the {\st{model}\af\em(what model?)} framework is trained to predict the difference between consecutive states rather than the states themselves. 
Given the feature matrix $\mathbf{R}$, the training target matrix $\mathbf{Y}$ specifies what the NGRC is trained to predict the future of. Specifically, we employ the NGRC as a `one-step-ahead integrator' like in Gauthier \textit{et al,}\cite{gauthier2021next}, training it to predict the difference between future consecutive states.
%{\dk \em(of the dynamical system from which the training data corresponds to rather than the states themselves)}. 
%{\af which is standard in NGRCing.\dk for me standard, not generally}
%{\af the problem here is that a reviewer may become suspicious as to whether this is standard or not based on how it is written and by specifying and citing a paper it may cause less suspicion}
%{\af\em(let's discuss this point on a call)}
Accounting for the warm-up phase, the `target matrix' is hence defined as
\begin{equation}
\mathbf{Y} = [\Delta \mathbf{x}_{\delta t}, \Delta \mathbf{x}_{\delta t + 1}, \ldots, \Delta \mathbf{x}_{T-1}] \in \mathbb{R}^{d \times (T - 1 - \delta t)},
\end{equation}
%{\af \st{\em(ok now I see the problem even clearer, there are $T-1-\delta t$ vectors of dimension $d$ in $Y$ but there are $T-\delta t$ vectors of dimension $Z$ in $R$, the number of vectors in $Y$ or $R$ needs to be adjusted and if $Y$ is adjusted then $\Delta x_{i}$ needs to be adjusted. i.e., currently $\mathbf{r}_{\delta t}$ maps to $\Delta\mathbf{x}_{\delta t}$, $\mathbf{r}_{\delta t+1}$ maps to $\Delta\mathbf{x}_{\delta t+1}$, ..., $\mathbf{r}_{T-1}$ maps to $\Delta\mathbf{x}_{T-1}$, but $\mathbf{r}_{T}$ is not mapped to anything ...)} \dk adjusted R to T-1}
%{\af\em(I'm guessing to keep the notion of a `one-step-ahead integrator' then we should alter $R$)}
where we define $\Delta \mathbf{x}_i = \mathbf{x}_{i+1} - \mathbf{x}_i$ which, by construction, enables the NGRC to act as a one-step-ahead integrator and, as a consequence, only $T-1-\delta t$ data points of $\mathbf{X}$ are used to train the NGRC.
%{\dk Explained before the equation so thinks thats fine}{\af \em(this sounds wrong, the difference in x at time step $t_i$ is the value of x at time step i+1 minus the value of x at time step i? it would be more appropriate to say, $\Delta \mathbf{x}_i = \mathbf{x}_{i} - \mathbf{x}_{i-1}$ for $i\geq1$. Also the the term `temporal difference' only makes sense if the temporal difference in [quantity Z] is specified )}
%{\dk Ya one can define it both ways its $x_i$ leads to $r_i$ that is mapped on $\Delta x_i$ that then leads to $x_{i+1}$, the indexing in the paper is based on that }.
%{\af\em(let's discuss this point on a call)}
%Each column of $\mathbf{Y}$ aligns with the corresponding feature vector in $\mathbf{R}$.
% {\af 
In other words,
% }
we train the NGRC to `learn' the mapping $\mathbf{r}_i \mapsto \Delta \mathbf{x}_i$. After training, 
% {\af
we obtain a
% }
predicted trajectory 
% {\af
on the attractor
% {\af\st{evolves according to}}
by iterating the following `NGRC system' forward through time,
\begin{equation}\label{eq:NGRC}
\mathbf{x}_{i+1} = \mathbf{x}_i + G(\mathbf{r}_i),
\end{equation}
where $G$ describes the trained output layer and maps feature vectors to the difference between 
% {\af\st{states}
successive predicted points on the trajectory.
% }
% {\af
To be more explicit, we use the known value of $\mathbf{x}_{i}$ at time $i=T$ to obtain a prediction of $\mathbf{x}_{i}$ at time $T+1$, and so on.
% }
%The discrete-time system defined in Eq.~\eqref{eq:NGRC} is what we refer to as the NGRC {\dk framework ?}.
%{\ms{[$Z = \sum_{p=1}^O \frac{(dk+p-1)!}{p!(dk-1)!}$} is the number of of features, the summands are the formula for combination with replacment. p=1 results in the number of features for the linear part, equivalenty the VAR model. For $p \geq 2$ nonlinear features are included, describing NVAR and NML. Intuitively, we have a set of $kd$ (linear) features and we want to create a new feature by multiplying $p$ features from the linear set. A feature from the linear set can appear more than once, hence with replacement and the order does not matter, thus we create \emph{unique} features.]}. 
%{\af\em(The training targets, i.e., the contents of the $Y$ matrix, need to be specified here, it is mentioned later in the multi-attractor task but its structure is also not specified)}
%In {\af\st{traditional reservoir computing applications} NGRC},
%{\af training consists of calculating only one output layer, $\mathbf{W}_{out}$, via ridge regression. This training method is used to find a desired map that minimizes the difference between the value of the final feature vector at a given point in time and the difference between the corresponding input data at the same time and the next point in time. The purpose of this is to arrive at the following equation, }
%\begin{equation}
%\mathbf{x}_{i+1}=\mathbf{x}_i+\mathbf{W}_{out} \mathbf{r}_{i+1},
%    \label{eq:differenceNG}
%\end{equation}
%where $x_{i}$ corresponds to predictions of the future of the input data used during training. 

\textbf{Training an NGRC+ERT system to perform attractor reconstruction: }
While standard NGRCs compute $G$ via ridge regression, the TreeDOX framework introduced by Giammarese \textit{et al.}~\cite{giammarese2024tree} employs Extremely Randomized Trees (ERT) to learn this mapping. 
% {\af\st{The}
We train using the same steps as above and use the ERT algorithm outlined in previous sections to obtain a predicted trajectory on the attractor by iterating the following 
% \st{resulting}
`NGRC+ERT system' 
% \st{is optimized to evolve}
forward through time, 
% \st{according to}
% }
\begin{equation}
\mathbf{x}_{i+1}=\mathbf{x}_i+\mathbf{ERT}( \mathbf{r}_{i}).
    \label{eq:differenceNG_ERT}
\end{equation}

\textbf{Training an NGRC+ERT system to perform multi-attractor reconstruction: }
We extend 
% {\af\st{this framework}
the work of Giammarese \textit{et al.}~\cite{giammarese2024tree}
% }
to multi-attractor reconstruction by integrating their TreeDOX framework with concepts from multifunctionality\cite{flynn2021multifunctionality} and parameter-aware learning\cite{kim2021teaching,kong2021machine}. This enables the NGRC+ERT system to 
% {\af\st{learn} 
reconstruct
% x}
the dynamics of $N$ distinct attractors with a single trained output layer. Each attractor $j$ is assigned a unique identifier $\theta_j \in \mathbb{N}$, which is multiplied with a scaling parameter $\gamma \in \mathbb{N}_{0}$ and added to every element of the attractor's corresponding feature vector, yielding the parametrized representation $\mathbf{r}'_{j,i} = \mathbf{r}_i + \gamma \theta_j$. 
The system operates in a multifunctional sense for $\gamma = 0$ and parameter-aware sense for $\gamma>0$.
For each
% {\af\st{each attractor time series}}
$\mathbf{X}_j$, the corresponding parametrized feature matrix $\mathbf{R}_j = [\mathbf{r}'_{j,\delta t}, \ldots, \mathbf{r}'_{j,T-1}]$ and target matrix $\mathbf{Y}_j$ are constructed as before. All $\mathbf{R}_j$ and $\mathbf{Y}_j$ matrices are concatenated into larger $\mathbf{R}_N$ and $\mathbf{Y}_N$ matrices for training, enabling the NGRC+ERT to learn the mapping $\mathbf{R}_j \mapsto \mathbf{Y}_j$ for each $j$ simultaneously.
% {\af\em(does it learn the mapping from matrix to matrix or from vector to vector, surely vector to vector to be consistent with how it is defined for a single attractor) \dk clear because concatination definition ? } for all attractors. 
For prediction, selecting a specific identifier $\theta_j$ enables the system to recall the corresponding attractor dynamics by iterating the following forward through time
\begin{equation}
\mathbf{x}_{i+1} = \mathbf{x}_i + \text{ERT}(\mathbf{r}'_{j,i}).
\label{eq:differenceNG_ERT_parameter}
\end{equation}
%The `two shot' component of our study is implemented as follows. Using the feature importance measure $\mathrm{I}$ provided within the ensemble of trained extremely randomized trees, we define a feature vector update rule that effectively acts as a filter in the feature space. Only those features whose measured importance exceeds a critical threshold, denoted by $\mathrm{I}_c$, remain active in the NGRC+ERT system. {\af Note, different values of $\mathrm{I}_c$ are used in different tasks according to the formulas specified in the main text.} For a given feature vector, $\mathbf{r}' \in \mathbb{R}^Z$, we construct the reduced representation $\tilde{\mathbf{r}}' = \Pi_{\mathrm{I}_c}(\mathbf{r}')$ through a projection operator that retains only the features $f$ where $\mathrm{I}_f > \mathrm{I}_c$, thus reducing the effective dimensionality from $Z$ to $\tilde{Z} = |\{f : \mathrm{I}_f > \mathrm{I}_c\}|$.{\ms $m$ to $Z$ for consistency} We then retrain $\mathbf{ERT}$ using these updated feature vectors $\tilde{\mathbf{r}}'$, where each retained element has demonstrated its significance by surpassing the importance cut-off $\mathrm{I}_c$, resulting in a new model $\widetilde{\mathbf{ERT}}$, for which the prediction of each attractor $j$ is performed by 

\textbf{Two-shot learning: }
We introduce a `two-shot' component in our experiments, 
% {\af
training the NGRC+ERT system and then updated it to produce a more optimal system.
% }
% {\af\st{that} 
Specifically, we use
% }{\af\st{leverages}}
the feature importance measure $\mathbf{FI}$ provided by the trained ERT ensemble to filter the feature space for an optimized feature configuration. Only features whose importance exceeds a critical threshold $\mathrm{I}_c$ 
% {\af
are allowed to
% }
remain active in the updated
% {\af
NGRC+ERT
% }
system. For a given feature vector $\mathbf{r}' \in \mathbb{R}^Z$, we construct the reduced representation $\tilde{\mathbf{r}}' = \Pi_{\mathrm{I}_c}(\mathbf{r}')$ through a projection operator that retains only features $f$ satisfying $\mathbf{FI}(f)> \mathrm{I}_c$ (Fig. \ref{fig:feature_importance_NEW} (b)), reducing the effective 
% {\af\st{dimensionality}
number of features
% }
from $Z$ to $\tilde{Z} = |\{f \mid \mathrm{I}_f > \mathrm{I}_c\}|$. 
%We then {\af\st{retrain an second ERT instance as the new systems output layer} train a new ERT} using these reduced feature vectors, yielding a refined model {that we denote by} $\widetilde{\text{ERT}}$. In the Results section, we analyze {\af\st{systems performance} the performance of the new NGRC+ERT system} using {\af the following} two feature importance thresholds $\mathrm{I}_c$. The first is an uniform threshold defined as $\mathrm{I}_c = 1/Z$, where $Z$ is the total number of features. The second is an exponential threshold defined as $\mathrm{I}_c = \max(\mathbf{FI}(f))/e$, where $e$ is Euler number. The attractor recall task is hence performed via,
We then train a new ERT using the retained feature vectors, yielding a refined 
% {\af\st{model}
ERT
% }
that we denote by $\widetilde{\text{ERT}}$. In the Results section, we analyze the performance of this two-shot approach using the following two feature importance thresholds $\mathrm{I}_c$. The first is a `uniform threshold', 
% {\af
similar to Giammarese \textit{et al.}\cite{giammarese2024tree},
% }
defined as $\mathrm{I}_c = 1/Z$, where $Z$ is the total number of features. The second is an exponential threshold defined as $\mathrm{I}_c = \max(\mathbf{FI}(f))/e$, where $e$ is Euler's number. The 
% {\af\st{attractor recall task}
reconstruction of attractor $j$
% }
is hence performed via,
\begin{equation}
\mathbf{x}_{i+1} = \mathbf{x}_i + \widetilde{\text{ERT}}(\tilde{\mathbf{r}}'_{j,i}),
\label{eq:differenceNG_ERT_parameter_optimized}
\end{equation}
% {\af 
and the same procedure is used to reconstruct the dynamics of each of the $N$ attractors.
% }

%{\af\em(should this Eq be quoted before the statement on computing system performance?) \dk yes, I see no issues here}

\begin{figure}[t]
  \centering
  \includegraphics[width=0.85\textwidth]{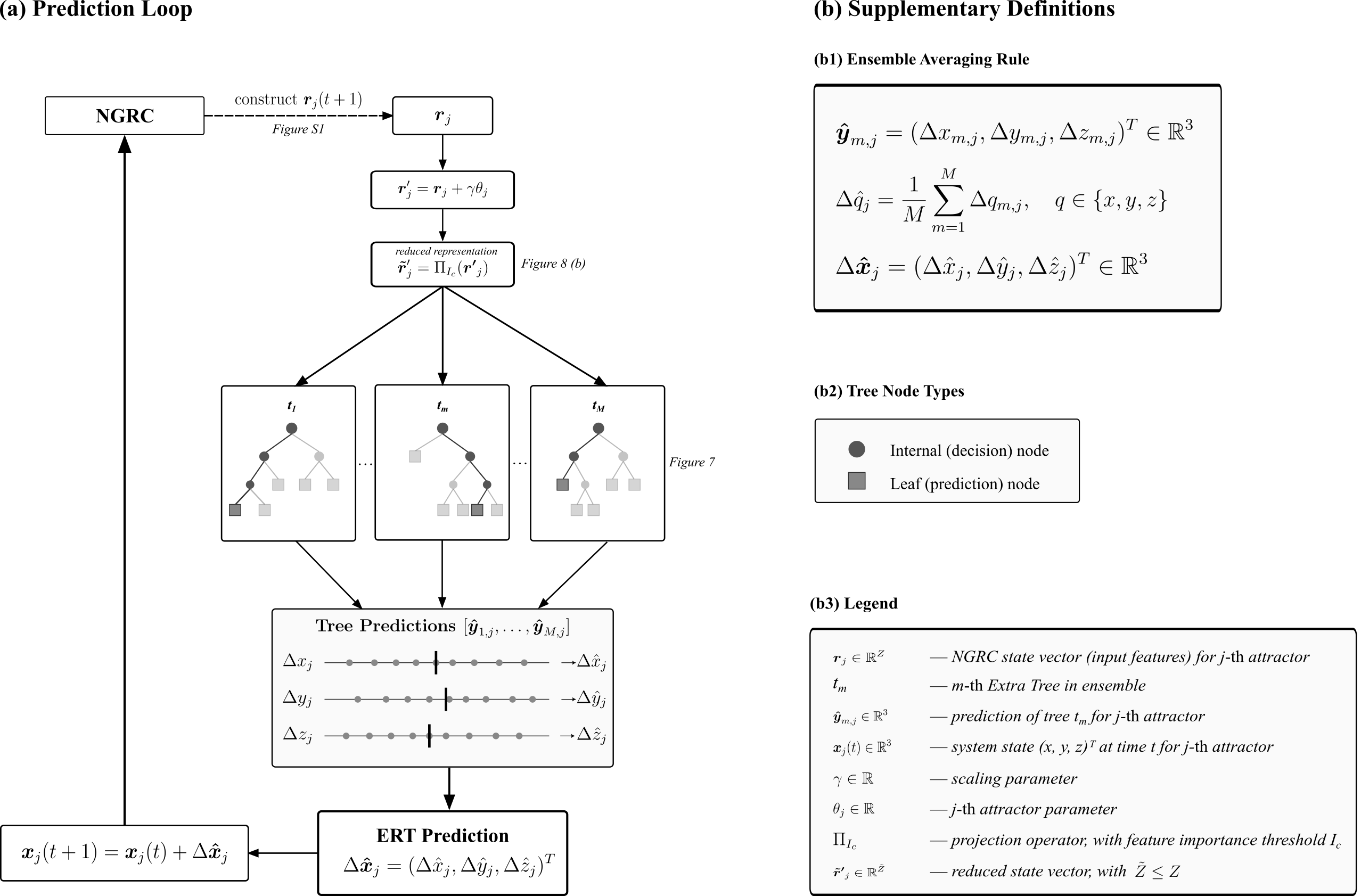}
  \caption{\textbf{Prediction loop of the NGRC+ERT system in a multi-attractor reconstruction setup} After training, the model predicts the next state of the $j$-th attractor by constructing a feature vector using the NGRC framework, augmented with the attractor-specific parameter $\theta_j$ scaled by $\gamma$. This vector is then projected onto a reduced feature space containing only the important features identified during training. Each tree in the ensemble receives the reduced vector and produces a prediction. The final output is obtained by averaging across all trees and adding the result to the current state.}
  \label{fig:prediction_loop}
\end{figure}

\subsection*{Description of tasks the NGRC+ERT system is trained to solve}

% \subsubsection*{Memory storage and recall}

% To test the accuracy of our system in terms of it's short-term predictive performance and long-term memory recall, we consider the task of attractor reconstruction, which is a memory storage and recall tasks, whereby we would like the NGRC+ERT system to provide accurate short-term predictions on how the state of a given system evolves on a given attractor while also reproducing the long-term statistical properties of the attractor. A hallmark of the brain is the ability to store and recall multiple different memories. We study the short-term and long-term predictions of our NGRC+ERT when trained to store and recall multiple different attractors without retraining. There have been many studies on this already, see multifunctionality papers. What complicates this task further is when the attractors from different attractors are overlapping in the input space. To this extent we consider three variations of this problem in the following tasks where we consider different combinations of attractors in different tasks:

We examine the performance of a NGRC+ERT system in different tasks, each of which involve training a NGRC+ERT system to reconstruct the dynamics of $N$ attractors. All of the attractors we consider are subsets of the state spaces of the various continuous-time dynamical systems listed in the Attractors section below. Training data, i.e., elements of the input time series $\mathbf{X}$ corresponding to each attractor, are obtained by integrating the state of a given dynamical system from a chosen initial condition up to time $(T+5000)dt$, using the 4$^{th}$-order Runge-Kutta method with integration time step $\Delta t$, and then discarding the transient of the first $5000$ time steps. From this we obtain a time series of length $T$ which describes a trajectory on a given attractor that we use as 
% {\af\st{training data}
$\mathbf{X}$.
% }
We continue to integrate the dynamical system for $P$ additional time step and reserve the corresponding time series of length $P$ as the `test', i.e., ground truth, for assessing the NGRC+ERT system's performance in predicting the dynamical system over a prediction length of $P$ time steps.
%\st{For a given dynamical system described in the following tasks that the NGRC+ERT system is trained to solve, we obtain a trajectory on the corresponding system's attractor by integrating the system equations forward in time using the 4th-order Runge-Kutta (RK4) method and discarding the transient of $1000$ time steps to obtain a time series of $T$ data points corresponding to the state of the system as it traverses the system's attractor.}

\noindent
\subsubsection*{Lorenz and Halvorsen task:}

%{\af\em(the lorenz attractor and the lorenz system are two different concepts that you have combined into one, which I have not seen done before. Please disassociate these concepts and do the same for the other attractors. Typically, we say we obtain a trajectory on the Lorenz attractor by integrating Eqs.~1-3 forward in time from initial condition, [], using the [4th-order Runge-Kutta] method with time step = [] and discarding the transient to obtain a time series of $T$ data points corresponding to the state of the system as it traverses the attractor.)}

We consider the Lorenz system described in Eqs.~\ref{eq:lorx}-\ref{eq:lorz} and the Halvorsen system described in Eqs.~\ref{eq:halx}-\ref{eq:halz}, and obtain trajectories on the corresponding Lorenz and Halvorsen attractors by integrating both systems using the 
% {\af\st{RK4 method}
4th-order Runge-Kutta (RK4) method
% }
with time step size $\Delta t = 0.02$ 
%{\dk \st{from initial conditions, $(x_{0}, y_{0}, z_{0})=(,,)$ and $(,,)$ respectively}}.
% {\af\st{After normalizing} 
We normalize
% }
both obtained trajectories to zero mean and unit standard deviation so that they overlap in state space 
% {\af 
to provide an additional level of difficulty; see Flynn \textit{et al.}~\cite{flynn2023seeing} and Kong \textit{et al.}~\cite{kong2024reservoir}.
% }
For all tests, we apply the attractor reconstruction metrics (specified later) over a prediction length of $P=15{,}000$ time steps.

\noindent
\subsubsection*{16-overlapping-attractor task:}

%For the 16‑overlapping‑attractor task we choose %{\af\em{two instances of each {\dk dynamical }system type} (it is not clear what this means)}%{\af\em(let's discuss this point on a call)}
For the 16-overlapping-attractor task, we generate time series from two realizations of trajectories on attractors from each of the eight dynamical systems described in the Attractors section by integrating each system from two different initial conditions for 20,000 time steps each. Each trajectory we obtain is normalized to have zero mean and unit standard deviation. To add an extra level of difficulty to the multi‑attractor task, we 
% {\af\st{randomly}}
rotate each 
% {\af\st{system}}
trajectory along every coordinate axis 
% {\af
so that all trajectories overlap with each other to varying extents.
% }
%{\af\em(are these attractors rotated so that they overlap, like in seeing double? overlap indeed provides an extra level of difficulty) (let's discuss this point on a call)}. Kong \textit{et al.}\cite{kong2024reservoir} demonstrated that classic multifunctional reservoir computers can scale to reconstruct {\af more and more} attractors {\af\st{from multiple dynamical systems}} simultaneously {\af by increasing the number of artificial neurons}. {\af\em(include picture on 'linear scaling law' OR remove reference to Kong and have a follow up paper where RC and NGRC are compared. Title; Different scaling laws apply to different types of multifunctional reservoir computers)} {\af\em{Here we show that the NGRC+ERT system can scale as well {\af by ... having a bigger tree? what is nice about NGRC+ERT as opposed to Kong is that he needs to add more features, i.e., more neurons, whereas NGRC+ERT does not, right?}, {\af\st{but} and do so} with the additional step of applying random orientations to the attractors - a variation that, to the best of our knowledge, has not yet been explored in the context of multifunctional reservoir computers.} (let's discuss this point on a call, I think Kong's attractors overlap too)}
Additionally, 
% {\af
to further illustrate the NGRC+ERT system's ability to reconstruct attractors from different dynamical systems,
% \st{we employ}
we add another level of difficulty by using
% }
different time step sizes for each system, specifically: $\Delta t = 0.005$ for Chen systems, $\Delta t = 0.02$ for Chua Circuit, Halvorsen, and Lorenz systems, $\Delta t = 0.08$ for Complex Butterfly, Windmi, and Rucklidge systems, and $\Delta t = 0.1$ for Roessler systems. %{\af\em(is there a reason why different $\Delta t$ are used?)\dk discussed in call, include here ? }
We apply the attractor reconstruction metrics, described later, over a prediction length of $P=15{,}000$ time steps across all tests. 
%{\af\em(what is `prediction length'? is this the $P$ that I defined? please also specify the $T$ for these)}
%{\af\em{(why not generate each attractor with the same time step?)}}

\subsubsection*{Multiple-overlapping-attractor task:}

%{\dk We investigate the optimized feature configuration obtained after training the NGRC+ERT system to reconstruct the dynamics of different numbers of attractors with the  exponential filtering scheme and the uniform filtering scheme .}
The task involves 4, 9, 16, 25, 36, and 49 overlapping attractors from different dynamical systems. 
The 
% {\af 
training data from the different
% }
attractors 
% {\af
we use in this task are obtained by
% \st{types are distributed by}
% }
cycling through the available dynamical systems (described in the Attractors subsection) 
% {\af\st{to}}
ensuring roughly equal representation 
% {\af
of each attractor.
% } 
Each trajectory is normalized to zero mean and unit standard deviation, and 
% {\af\st{randomly}}
rotated along every coordinate axis 
% {\af
for the same reasons as before.
% } 
The time step $\Delta t$ for each system is the same as specified in the 16-overlapping-attractor task description.

%The multiple-overlapping-attractor tasks involves 4, 9, 16, 25, 36, and 49 overlapping attractors from different dynamical systems. The task setup follows the 16-overlapping-attractor task. {\af\em{The types of dynamical systems used are randomly selected from those described in the Attractors subsection.} (what does this mean?)} The obtained trajectories are normalized to zero mean, unit standard deviation and {\af\em{randomly rotated in phase space}(let's discuss this point on a call)}.
%{\af{randomly rotated}??? how are they randomly rotated? or is there some principle involved in how you choose to orient them?}
%Each chosen system has a specific time step, $\Delta t = 0.005$ for Chen systems, $\Delta t = 0.02$ for the Chua circuit and the Halvorsen, and Lorenz systems, $\Delta t = 0.08$ for Complex Butterfly, Windmi, and Rucklidge systems, and $\Delta t = 0.1$ for Roessler systems. {\af can just say $\Delta t$ for each system is the same as specified in the 16-overlapping-attractor task description.}

\subsection*{Attractor reconstruction metrics}
We now describe the two metrics we use
to assess how accurate the NGRC+ERT system learns the \textbf{two attractor tasks} in terms of long-term and short-term behavior. \\
\textbf{Long-term behavior:}
We define the `long-term measurement error' in terms of how the largest Lyapunov exponent~\cite{rosenstein1993practical} (LY) $\lambda$ and correlation dimension~\cite{grassberger2004measuring} (CD) $\nu$ of a given reconstructed attractor compares to its corresponding ground truth across all $N$ analyzed attractors here.
More specifically, we compute the absolute relative error of the LY and CD of a given reconstructed attractor as,

\begin{equation*}
    \Delta \lambda = \frac{|\lambda^{\text{test}} - \lambda^{\text{pred}}|}{\lambda^{\text{test}}}, \qquad
    \Delta \nu = \frac{|\nu^{\text{test}} - \nu^{\text{pred}}|}{\nu^{\text{test}}},
\end{equation*}
where \textit{test} correspond to the ground truth  and \textit{pred} to the reconstructed attractor over the prediction interval defined by $P$.
We define the long-term measurement error (LTME) as,
\begin{equation}
    \text{Long-term measurement error} = \frac{1}{2N} \sum_{i=1}^N (\Delta \lambda_i + \Delta \nu_i).
    \label{eq:Long_term_measurement_error}
\end{equation}
We consider the NGRC+ERT system to reconstruct the `climate' and performs accurate long-term memory recall of both attractors in the sense of Lu \textit{et al.}\cite{lu2018attractor} if the $\text{LTME} \leq 0.1$ which was found empirically. If the $\text{LTME} > 0.1$ the reconstructions begin to break down.
%{\af [cite papers where this metric has been used before?]}
\\
\textbf{Short-term behavior:}
We define the `short-term measurement error' in terms of the `mean forecast horizon' (MFH) across all $N$ attractors. More specifically, we compute the `forecast horizon' (FH) of a given reconstructed attractor and its corresponding ground truth both over the interval defined by $P$ as 
%{\af\st{To evaluate the short-term behavior of the predicted trajectory, we use the forecast horizon (FH) $\tau$. Given a predicted time series $\mathbf{X}{}_{\text{pred}}$ and its corresponding test time series $\mathbf{X}_{\text{test}}$,}}
%{\af\em(these do not describe time series, they describe a point in the time series ..., time series in this sense is a matrix OR x(t) defined for t = 1, 2, ..., T?)\dk good catch}
%{\af\st{the forecast horizon measures}}
the number of time steps for which the prediction time series, $\mathbf{X}_{\text{pred}}$, remains within a threshold $\mathbf{\epsilon} \in \mathbb{R}^d$ of the test time series $\mathbf{X}_{\text{test}}$ across all $d$ variables:
\begin{equation*}
    \tau = \arg\max_t (\mid \mathbf{x}_{train}(t) - \mathbf{x}_{test}(t)| < \mathbf{\epsilon})  
\end{equation*}
The threshold vector $\mathbf{\epsilon}$ is defined element-wise as a fraction of the difference between the maximum and minimum values of $\mathbf{x}_{\text{test}}(t)$ in each variable. In our case we found the following
\begin{equation*}
    \boldsymbol{\epsilon} = 0.15\cdot(\max\mathbf{x}_{test}(t) - \min\mathbf{x}_{test}(t)),
\end{equation*}
to be sufficient, where the $\max$ and $\min$ operators are applied separately to each variable. To quantify the short-term behavior in the multi-attractor reconstruction setup, we compute the mean forecast horizon across all $N$ attractors:
\begin{equation}
    \text{Mean Forecast Horizon} = \frac{1}{N} \sum_{i=1}^N \tau_i',
    \label{mean_forecast_horizon}
\end{equation}
where each forecast horizon is expressed in Lyapunov time $\tau' = \tau \cdot \lambda^{\text{test}} \cdot \Delta t$. This normalization allows direct comparison across dynamical systems with different time scales ($\Delta t$) and largest Lyapunov exponents ($\lambda^{\text{test}}$).

For the \textbf{16-overlapping-attractor task}, we compare the LY, CD, and compute the FH of each reconstructed attractor individually against the corresponding ground truth over the prediction length $P$.

\subsection*{Attractors}

%{\dk \em (We obtain trajectories on the attractors from each of the systems specified below by integrating these systems forward through time using the 4th-order Runge-Kutta (RK4) method with time step size $\Delta t$ whose values were specified in the section above.) discribed above}

\subsubsection*{Chen System}
\begin{align}
\dot{x} &= a(y - x ) \label{eq:chenx}\\
\dot{y} &= (c-a)x - xz + cy \label{eq:cheny} \\
\dot{z} &= xy - bz \label{eq:chenz}
\end{align}
\noindent
with parameters $a = 35$, $b = 3$, $c = 28$ for the 16-overlapping-attractor task and an additional parametrization of $a = 33$, $b = 3$, $c = 28$ for the multiple-overlapping-attractor task.

\subsubsection*{Chua Circuit}
\begin{align}
\dot{x} &= \alpha(y - x + bx + 0.5 (a-b) ((\left|x\right| +1) -(\left|x\right| -1))) \label{eq:chuax}\\
\dot{y} &= x-y+z \label{eq:chuay}\\
\dot{z} &= -\beta y \label{eq:chuaz}
\end{align}
\noindent
with parameters $\alpha=9$, $\beta = 100/7$, $a = 8/7$, $b = 5/7$ for the two-attractor task and the 16-overlapping-attractor task and an additional parametrization of $\alpha=8.5$, $\beta = 100/7$, $a = 8/7$, $b = 5/7$ for the multiple-overlapping-attractor task. 

\subsubsection*{Complex Butterfly System}
\begin{align}
\dot{x} &= a(y-x) \label{eq:cbx}\\
\dot{y} &= -z \frac{x}{\left|x\right| } \label{eq:cby}\\
\dot{z} &= \left|x\right| -1 \label{eq:cbz}
\end{align}
\noindent
with parameter $a = 0.55$ across all tasks.

\subsubsection*{Halvorsen System}
\begin{align}
\dot{x} &= -\sigma x -4y -4z -y^2 \label{eq:halx} \\
\dot{y} &= -\sigma y -4z -4x -z^2 \label{eq:haly} \\
\dot{z} &= -\sigma z -4x -4y -x^2 \label{eq:halz}
\end{align}
\noindent
with parameter $\sigma=1.27$ for the two-attractor task and the 16-overlapping-attractor task and an additional parametrization of $\sigma=1.7$ for the multiple-overlapping-attractor task.

\subsubsection*{Lorenz System}
\begin{align}
\dot{x} &= \sigma(y - x) \label{eq:lorx} \\
\dot{y} &= x(\rho - z) - y \label{eq:lory} \\
\dot{z} &= xy - \beta z \label{eq:lorz}
\end{align}
\noindent
with parameters $\sigma = 10$, $\rho = 28$, $\beta = 8/3$ for the two-attractor task and the 16-overlapping-attractor task and an additional parametrization of $\sigma = 10$, $\rho = 350$, $\beta = 8/3$ for the multiple-overlapping-attractor task.

\subsubsection*{Roessler System}
\begin{align}
\dot{x} &= -y - z \label{eq:rx}\\
\dot{y} &= x + a y \label{eq:ry}\\
\dot{z} &= b + z(x - c)\label{eq:rz}
\end{align}
\noindent
with parameters $a = 0.2$, $b = 0.2$, $c = 5.7$ for the two-attractor task and the 16-overlapping-attractor task.

\subsubsection*{Rucklidge System}
\begin{align}
\dot{x} &= -\kappa x + \lambda y - yz \label{eq:ruckx}\\
\dot{y} &= x \label{eq:rucky}\\
\dot{z} &= -z + y^2 \label{eq:ruckz}
\end{align}
\noindent
with parameters $\kappa = 2$, $\lambda = 6.7$ for the two-attractor task and for the multiple-overlapping-attractor task.

\subsubsection*{Windmi System}
\begin{align}
\dot{x} &= y \label{eq:wx}\\
\dot{y} &= z \label{eq:wy}\\
\dot{z} &= -az - y + b - e^x \label{eq:wz}
\end{align}
\noindent
with parameters $a = 0.7$, $b = 2.5$ for the two-attractor task and the 16-overlapping-attractor task and an additional parametrization of $a = 0.7$, $b = 1.75$ for the multiple-overlapping-attractor task.

\bibliography{sample}

\section*{Acknowledgements}
D. K. gratefully acknowledges the funding provided by Allianz Global Investors (AGI).

\noindent We dedicate this work to the late Erik Bollt who pursued similar research and pioneered the field. May he rest in peace.

%as presented in this work. We would have liked to  work jointly  on further NGRC / ERT projects in the future and deeply regret that this won't happen.}

\section*{Author contributions statement}

D.K. and C.R. conceived the study. M.S. and D.K. developed the methodology and conducted the experimental studies. A.F. contributed to the interpretation of results. D.K., M.S. and A.F. prepared the manuscript. All authors edited the manuscript. C.R. initiated and supervised the project.

\newpage
\setcounter{figure}{0}
\setcounter{table}{0}
\setcounter{equation}{0}
\renewcommand{\thefigure}{S\arabic{figure}}
\renewcommand{\thetable}{S\arabic{table}}
\renewcommand{\theequation}{S\arabic{equation}}
\renewcommand{\thesection}{S\arabic{section}}
\setcounter{section}{0}

\section*{Supplementary Note: Lorenz and Halvorsen task}

\begin{figure}[t]
  \centering
  \includegraphics[width=0.80\textwidth]{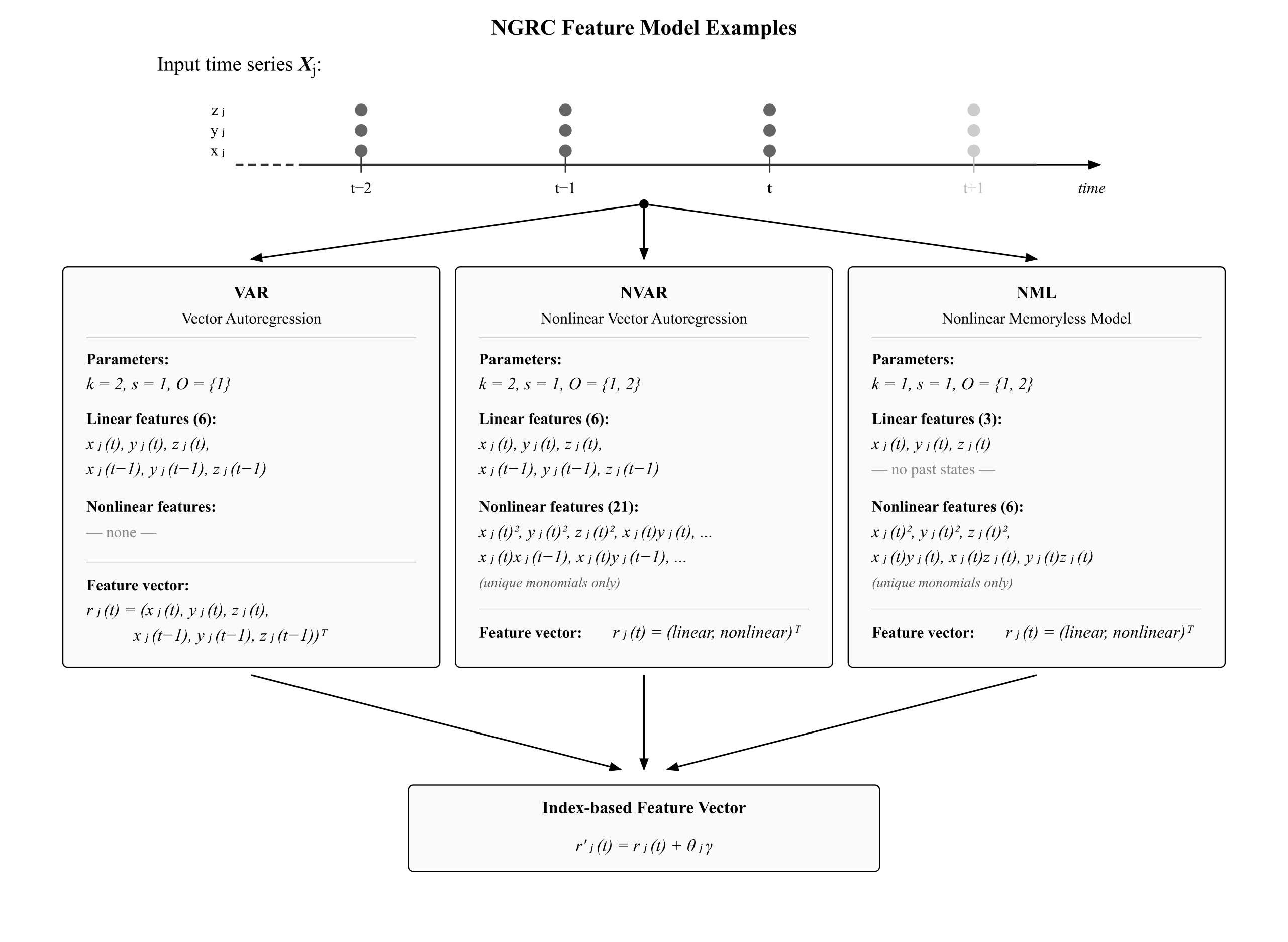}
  \caption{\textbf{Schematic of feature construction in NGRC for three feature model types.} Each model takes a time series as input. The VAR model constructs only linear features from past observations. The NVAR model extends this by adding nonlinear features, computed as unique higher-order monomials of the linear features. The NML model constructs nonlinear features using only the most recent observation. For all model types, each feature vector is augmented with an attractor-specific parameter (unique identifier) $\theta_j$ scaled by $\gamma$.}
  \label{fig:ngrc_diagram}
\end{figure}

\begin{figure}[t]
    \centering
    \includegraphics[width=\linewidth]{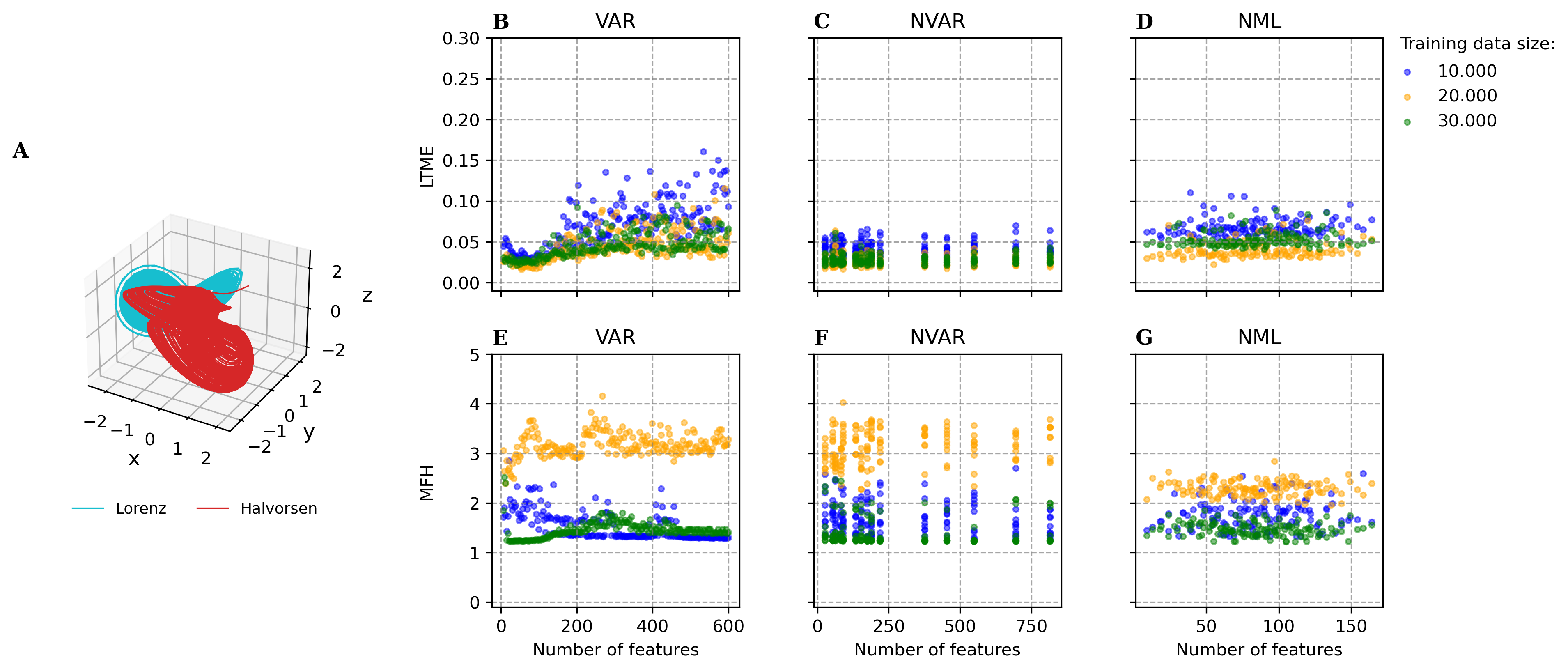}
    \caption{$\textbf{NGRC feature models applied on the Lorenz and Halvorsen task.}$ Each data point represents the mean over 10 random initializations of ERT instances. \textbf{A} Overlapping state space trajectories of the Lorenz and Halvorsen attractors after normalization. \textbf{B} Long-term measurement error for the VAR feature model. For a few features, a clear minimum is observed across all training set sizes, while the error increases when many features are included. With larger training sizes, the error becomes less sensitive to the number of features. \textbf{E} Mean forecast horizon for the VAR feature model. The training size has a substantial impact on short-term behavior. Across the tested hyperparameter configurations, the intermediate training size yields the highest forecast horizon.  \textbf{C} Long-term measurement error for the NVAR feature model. The error remains relatively constant across the number of features for all training sizes, with only minor differences between them. \textbf{F} Mean forecast horizon for the NVAR feature model. The influence of training size is similar to that observed for the VAR feature model, with the intermediate training size producing the highest forecast horizon. \textbf{D} Long-term measurement error for the NML feature model. Similar behavior as in the NVAR feature model is observed. The difference in training sizes is more pronounced. \textbf{G} Mean forecast horizon for the NML feature model. The influence of the training size is consistent with the other two models, but the forecast horizon is generally lower across all hyperparameter configurations.
    }
    \label{fig:Lorenz and Halvorsen ground}
\end{figure}

In the Results section, we present the main results from the following analysis of our NGRC+ERT system on the overlapping Lorenz and Halvorsen task. 
The aim of this task is to produce a NGRC+ERT system that combines its short-term and long-term memory processing capabilities to accurately reconstruct how the state of each different system evolves on the respective attractors (Lorenz and Halvorsen) with the added complication that the training data from these attractors overlap, i.e., share common points in state space (Figure~\ref{fig:Lorenz and Halvorsen ground} \textbf{A}). For more information on this complication see Flynn \textit{et al.}\cite{flynn2023seeing}. 
Here, we compare  the three distinct feature model types derivable from a NGRC setup illustrated in Figure \ref{fig:ngrc_diagram}. 
We examine how the VAR, NVAR and NML feature model types, and their chosen parameters $(k, s, O)$, affect the accuracy of the short-term and long-term predictions made by the NGRC+ERT system on the Lorenz and Halverson task.
To quantify the short-term prediction accuracy of our NGRC+ERT system, we use the mean forecast horizon (MFH) of both predicted attractors, as described in Methods.
%{\af The forecast horizon is the time} {\af\st{measuring the time steps until each}} {\af where the} prediction {\af first} deviates by 15 $\%$ to the ground truth in terms of the Lyapunov time of the attractor. {\af\em(mention what Lyapunov time is)} {The larger the mean forecast horizon, the better the short-term prediction of both attractors.}
The long-term prediction accuracy of our NGRC+ERT system on the two attractors is characterized by the average relative error over 10 random initializations of ERT instances based on the difference between (i) the largest Lyapunov exponent (LY) and (ii) the correlation dimension (CD) of a given attractor and the corresponding reconstructed attractor using the long-term measurement error (LTME) defined in detail in Methods.
%The long-term prediction accuracy of our NGRC+ERT system is {\af\st{characterized by the average relative error} based on the difference} between (i) the largest Lyapunov exponent (LY) and (ii) the correlation dimension (CD) of a given attractor and the corresponding reconstructed attractor. {\af We compute the average of the four quantities (differences between LY and CD for both attractors) and consider the resulting quantity as the} long-term measurement error{\af\st{ in the following}}. {\af \em(where does the `relative' aspect come in?)}{\af The smaller the long-term measurement error, the better the long-term predictions of both attractors.}{\af When this error is small we consider the NGRC+ERT system to successfully reconstruct the `climate' of both attractors in the sense of Lu {et al.}\cite{lu2018attractor}} 
Table~\ref{tab: parameter config NGRC_} details the hyperparameter configurations tested, with $\mathcal{D}(O)$ representing the set from which combinations of monomial orders are defined.

\begin{table}[t]
    \centering
    \begin{tabular}{|l|c|c|c|}
        \hline
        \textbf{Feature Model Type} & \textbf{$k$} & \textbf{$s$} & \textbf{$\mathcal{D}(O)$} \\
        \hline
        VAR & $2, 3, \ldots, 200$ & 1 & $\{1\}$ \\
        NVAR & $2, 3, 4, 5, 6$ & $1, 2, \ldots, 10$ & $\{1, 2, 3, 4\}$ \\
        NML & 1 & 1 & $\{1, 2, 3, 4, 5, 6, 7, 8\}$ \\
        \hline
    \end{tabular}
    \caption{Hyperparameter configuration of the NGRC feature models}
    \label{tab: parameter config NGRC_}
\end{table}

\begin{figure}[t]
    \centering
    \includegraphics[width=\linewidth]{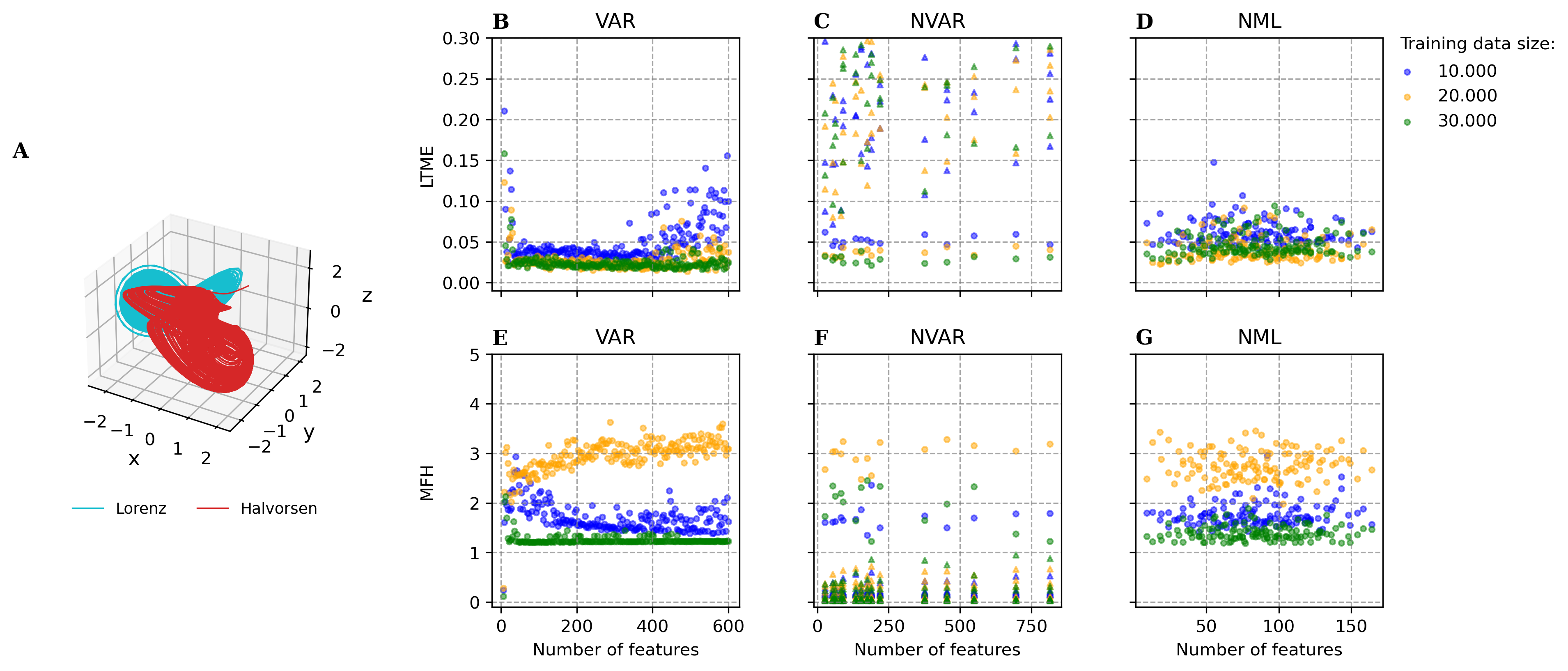}
    \caption{$\textbf{NGRC feature models on the Lorenz and Halvorsen task with the uniform scheme applied.}$ Each data point represents the mean over ten random initialized ERT instances. \textbf{A} Overlapping state space trajectories of the Lorenz and Halvorsen attractors after normalization. \textbf{B} For a sufficiently larger number of features, the long-term measurement error is comparable to the minimum observed in (Figure \ref{fig:Lorenz and Halvorsen ground}, \textbf{A}). Larger training sizes exhibit greater robustness with respect to the error when increasing the number of features. \textbf{B} For the NVAR feature model, similar error values are observed for $s = 1$ (circles), whereas for $s > 1$ (triangles) the error increases significantly. \textbf{C} The NML feature model with FI demonstrated more robustness across different training sizes. \textbf{E} The forecast horizon exhibits a similar trend to that observed without FI. \textbf{F} For $s > 1$, the forecast horizon tends to vanish, suggesting that larger temporal spacing between past consecutive trajectory points in the features is detrimental for short term performance. \textbf{G} For the NML feature model, the short-term prediction accuracy remains similar to that observed without FI.}
    \label{fig:Lorenz and Halvorsen with FI}
\end{figure}

\noindent
In Figure~\ref{fig:Lorenz and Halvorsen ground} we evaluate the short-term and long-term prediction accuracy among the three feature models VAR (Figures~\ref{fig:Lorenz and Halvorsen ground} \textbf{B, E}), NVAR (Figures~\ref{fig:Lorenz and Halvorsen ground} \textbf{C, F}), and NML (Figures~\ref{fig:Lorenz and Halvorsen ground} \textbf{D, G}) against hyperparameter configurations of each feature model. We also investigate the influence of varying training sizes $T \in \{10{,}000, 20{,}000, 30{,}000\}$. Each attractor is predicted for $P = 15{,}000$ steps. The results show that with the intermediate training size $T=20{,}000$ per attractor, the feature models yield the best performance both in the short-term and in the long-term prediction accuracy (Figures~\ref{fig:Lorenz and Halvorsen ground} \textbf{B-G}). Initially, increasing the training size from $10{,}000$ to $20{,}000$ leads to consistent improvement, regardless of the number of features or whether the feature model employs linear, nonlinear, or mixed features. However, performance degrades for $T = 30{,}000$, suggesting that beyond a certain threshold, additional data may impair the learning process. Models incorporating nonlinear features demonstrate reduced sensitivity to the number of features, whereas the VAR feature model exhibits a clear optimum where a small to intermediate number of features yields the best performance. 
%Too few features result in poor splits due to limited available information. 
In Figures.~\ref{fig:Lorenz and Halvorsen ground} \textbf{B, E} we observe a positive correlation between the number of  features and long-term measurement error across all training sizes in the VAR feature model, indicating that many features adversely affect performance. In particular, the mean forecast horizon reaches its minimum at the largest training size ($T = 30{,}000$) across all models. The observation that the appropriate selection of training data benefits both short-term and long-term prediction accuracy raises important questions about the underlying mechanisms. Although larger training datasets provide more information from which models can learn, the reason for performance degradation beyond certain amounts of data remains unclear. Previous research has shown that excessive data can induce instabilities in NGRC employing ridge regression~\cite{zhang2025more}. Our similar findings suggest that this phenomenon might be independent of the specific regression method used and may thus be a property of the feature model.
%Furthermore, the VAR feature model exhibits a distinct minimum for long-term measurement error and a corresponding maximum for short-term prediction accuracy, indicating the existence of an optimal time-delay feature configuration for this task.
In Figure~\ref{fig:Lorenz and Halvorsen with FI} we show the measurements for the three feature models when the uniform feature importance filtering scheme is applied. We take the same hyperparameter configurations used in Figure~\ref{fig:Lorenz and Halvorsen ground} and evaluate the predictive performance of each hyperparameter configuration when the uniform feature importance filtering scheme is switched on. We find that the long-term prediction accuracy improves within the VAR and NML feature models. Figure~\ref{fig:Lorenz and Halvorsen with FI} \textbf{B, E} show the VAR feature model exhibits the highest mean improvement in long-term prediction accuracy: on average $9\%$ for $10{,}000$ training steps, $15\%$ for $20{,}000$ training steps and $29\%$ for $30{,}000$ training steps across all analyzed hyperparameter configurations, calculated relative to the hyperparameter configurations where no feature importance scheme was applied. Although, in Figure~\ref{fig:Lorenz and Halvorsen with FI} \textbf{D, G}, the NML feature model showed partial increases in long-term prediction accuracy , $8\%$ for $10{,}000$ training steps, $-2\%$ for $20{,}000$ training steps, and $11\%$ for $30{,}000$ training steps, the performance of the NVAR feature model decreased substantially, particularly for configurations with $s>1$. For short-term prediction accuracy, we observed that in the VAR feature model, the forecast horizon increased on average by $13\%$ for $10{,}000$ training steps, while decreasing by $-7\%$ for $20{,}000$ training steps and by $-13\%$ for $30{,}000$ training steps. For the NML feature model, we observe an increase of  $4\%$ for $10{,}000$ training steps, $21\%$ for $20{,}000$ training steps, and a decrease by $-4\%$ for $30{,}000$ training steps. The results suggest than particularly for the VAR feature model the uniform feature importance scheme approach improves attractor reconstruction across certain hyperparameter configurations significantly (Figure \ref{fig:Lorenz and Halvorsen with FI} \textbf{B}).

\begin{figure}[t]
    \centering
    \includegraphics[width=\linewidth]{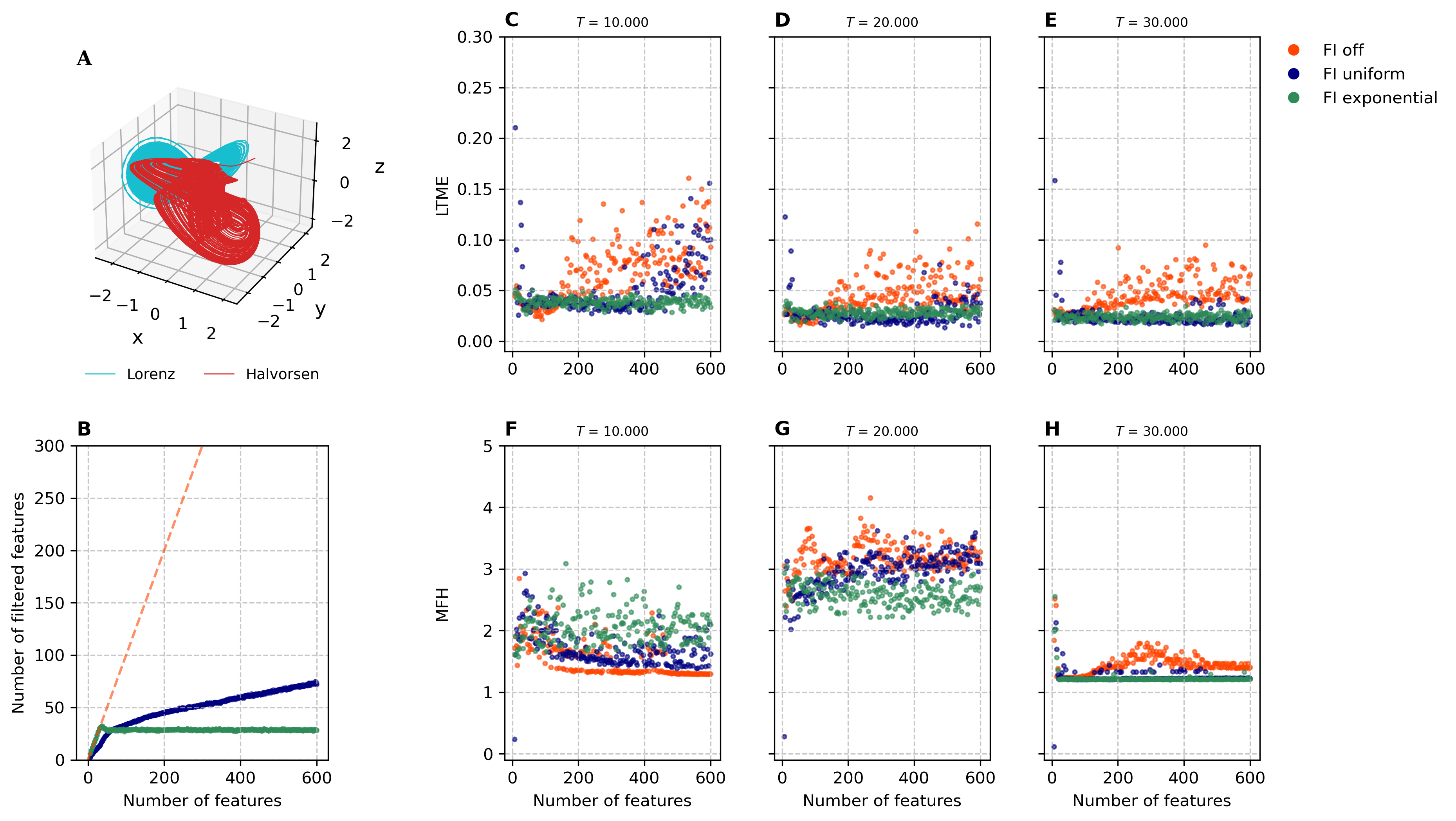}
    \caption{\textbf{VAR feature model on the Lorenz and Halvorsen task.} \textbf{A} Overlapping state space trajectories of the Lorenz and Halvorsen attractors after normalization. \textbf{B} Comparison of feature importance schemes: the uniform scheme shows linear growth in filtered features with increasing number of features, while the exponential scheme approaches a constant number of retained features. The exponential scheme performs minimal filtering for small initial numbers of features, whereas the uniform scheme removes more features, resulting in inferior performance (see Figure~\ref{fig:Relative improvement}). \textbf{C}-\textbf{E} Long-term measurement error with FI versus no FI (base model) applied. Both uniform and exponential scheme approaches achieve comparable error compared to the base model when the best performing features are selected. For smaller training sizes, the error in the uniform scheme increases with additional features while errors in the exponential scheme remain relatively constant, indicating that the model performance with limited training data suffers from large number of features. \textbf{F}-\textbf{H} Mean forecast horizon improves with FI for smaller training sizes, whereas for larger training sizes the forecast horizon shows slight degradation in short-term prediction accuracy. FI enhances long-term dynamics while marginally affecting short-term performance.
    }
    \label{fig:combined_analysis_all}
\end{figure}

\begin{figure}[t]
    \centering
    \includegraphics[width=\linewidth]{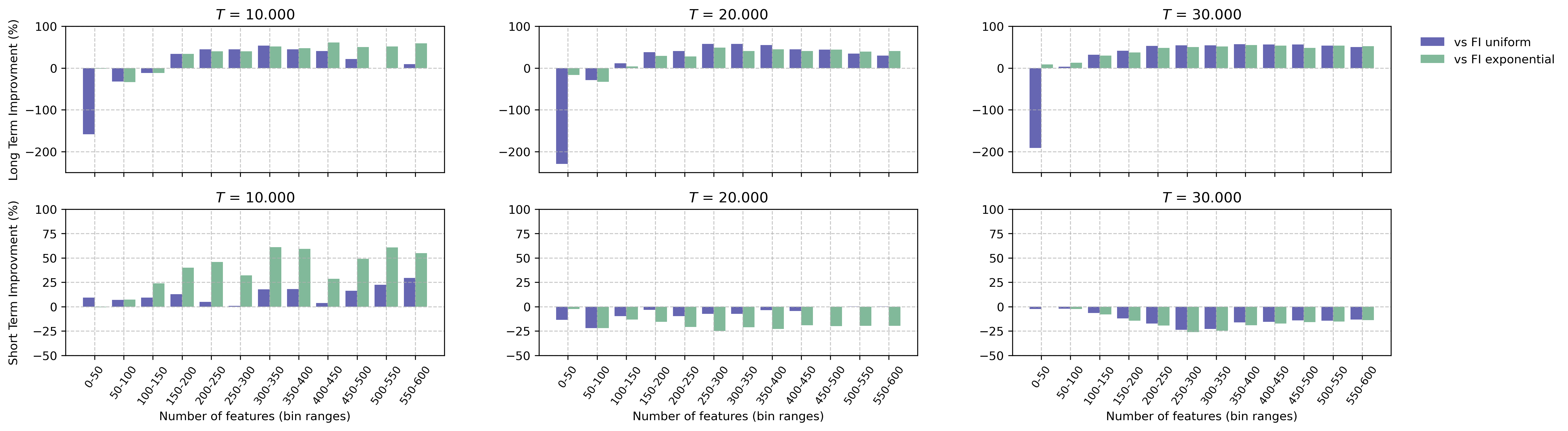}
    \caption{\textbf{Relative improvement of the uniform and exponential scheme against the base VAR feature model without FI for the Lorenz and Halvorsen task}. Top row displays the long-term metrics: For few features, the uniform scheme performs worse than the base model (see Figure \ref{fig:combined_analysis_all}, \textbf{A}) for less than 50 features, whereas the exponential scheme yields comparable errors. The feature importance schemes improve model performance for both approaches once the number of features is sufficiently large. For small training sizes, the exponential scheme achieves larger improvements. Bottom row displays the short-term metics relative to the VAR without FI. For a training size $T=10{,}000$, both the uniform and exponential scheme improve the forecast horizon, whereas for the larger training sizes, the base model outperforms both schemes marginally.}
    \label{fig:Relative improvement}
\end{figure}

In Figure~\ref{fig:combined_analysis_all}, we test the exponential feature importance filtering scheme approach introduced for the VAR feature model and evaluate the accuracy of the short-term as well as the long-term predictions. We find that applying the exponential scheme, the accuracy of the attractor reconstructions improves for the VAR feature model by $33\%$ for $10{,}000$ training steps, $26\%$ for $20{,}000$ training steps, and $42\%$ for $30{,}000$ training steps averaged across all analyzed configurations, calculated relative to the hyperparameter configurations without a feature importance scheme. For smaller training datasets, the forecast horizon improves when applying a feature importance scheme, with the exponential scheme showing particularly consistent results across all hyperparameter configurations. However, increasing the size of the training data results in slightly worse short-term prediction accuracy. We observed that the forecast horizon increased on average by $39\%$ for $10{,}000$ training steps, while decreasing by $-19\%$ for $20{,}000$ training steps and by $-15\%$ for $30{,}000$ training steps across all hyperparameter configurations compared to those where no feature importance scheme is applied. Although applying a feature importance scheme appears to favor improvements in attractor reconstruction, its improvements in the short-term prediction accuracy are observed solely for the smallest training size. Consequently, the introduction of the exponential scheme for ERT in the combination with the VAR feature model enables consistently high performance across the analyzed hyperparameter configurations for Lorenz and Halvorsen task, drastically reducing the need for hyperparameter optimization. We compare both FI schemes by measuring their relative improvement compared to when no FI scheme is applied, examining performance within specific feature number segments in Figure \ref{fig:Relative improvement}.

\begin{figure}[t]
    \centering
    \includegraphics[width=0.75\linewidth]{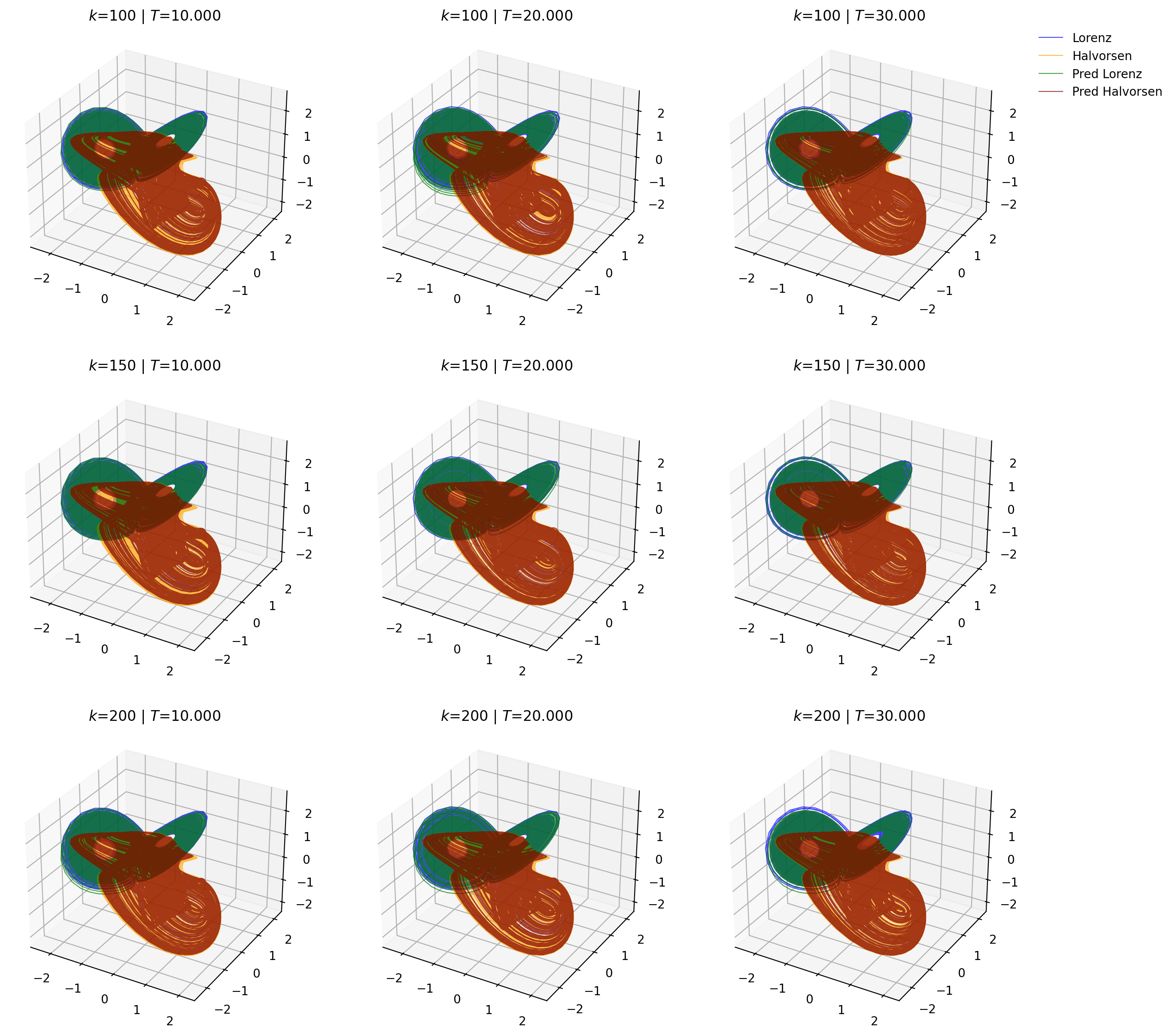}
    \caption{Prediction for the Lorenz and Halvorsen task for different $k$ values and training sizes $T$, using VAR feature model and the exponential scheme.}
    \label{fig:Lorenz + Halvorsen predictions FI exponential}
\end{figure}

% \newpage
\section*{Supplementary Note: Chua and Halvorsen task}

As an additional two attractor task, we consider reconstructing chaotic attractors from the Chua and Halvorsen systems and integrate both with a time step size of $dt = 0.05$. After normalizing both obtained trajectories to zero mean and unit standard deviation, we overlay them in state space. For all tests, we apply the attractor reconstruction metrics over a prediction length $P=15{,}000$ time steps.

\begin{figure}[t]
    \centering
    \includegraphics[width=\linewidth]{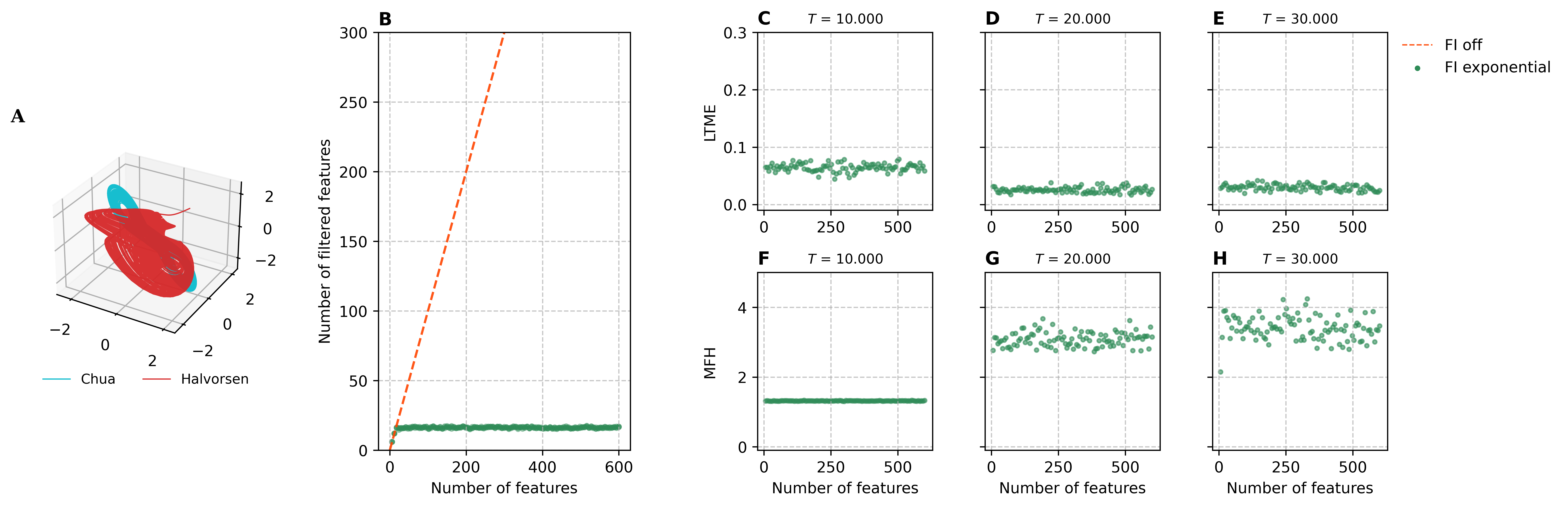}
    \caption{\textbf{Performance metrics of the exponential scheme on the Chua and Halvorsen task}. Each data point represents the mean over ten random initialized ERT instances. \textbf{A} Chua and Halvorsen overlapping in state space, normalized to mean 0 and standard deviation 1, with scaling parameter $\gamma = 5$. \textbf{B} The number of input features versus the number of filtered features for the VAR feature model. There is no significant filtering for few features observed. Beyond $\sim 50$ features, the number of filtered features plateaus and the number of filtered features remains at a value of 16. The Long-term measurement error is displayed in \textbf{C}-\textbf{E} and the mean forecast horizon is displayed in \textbf{F}-\textbf{H}.}
    \label{fig:Chua + Halvorsen combined}
\end{figure}
\begin{figure}[t]
    \centering
    \includegraphics[width=0.70\linewidth]{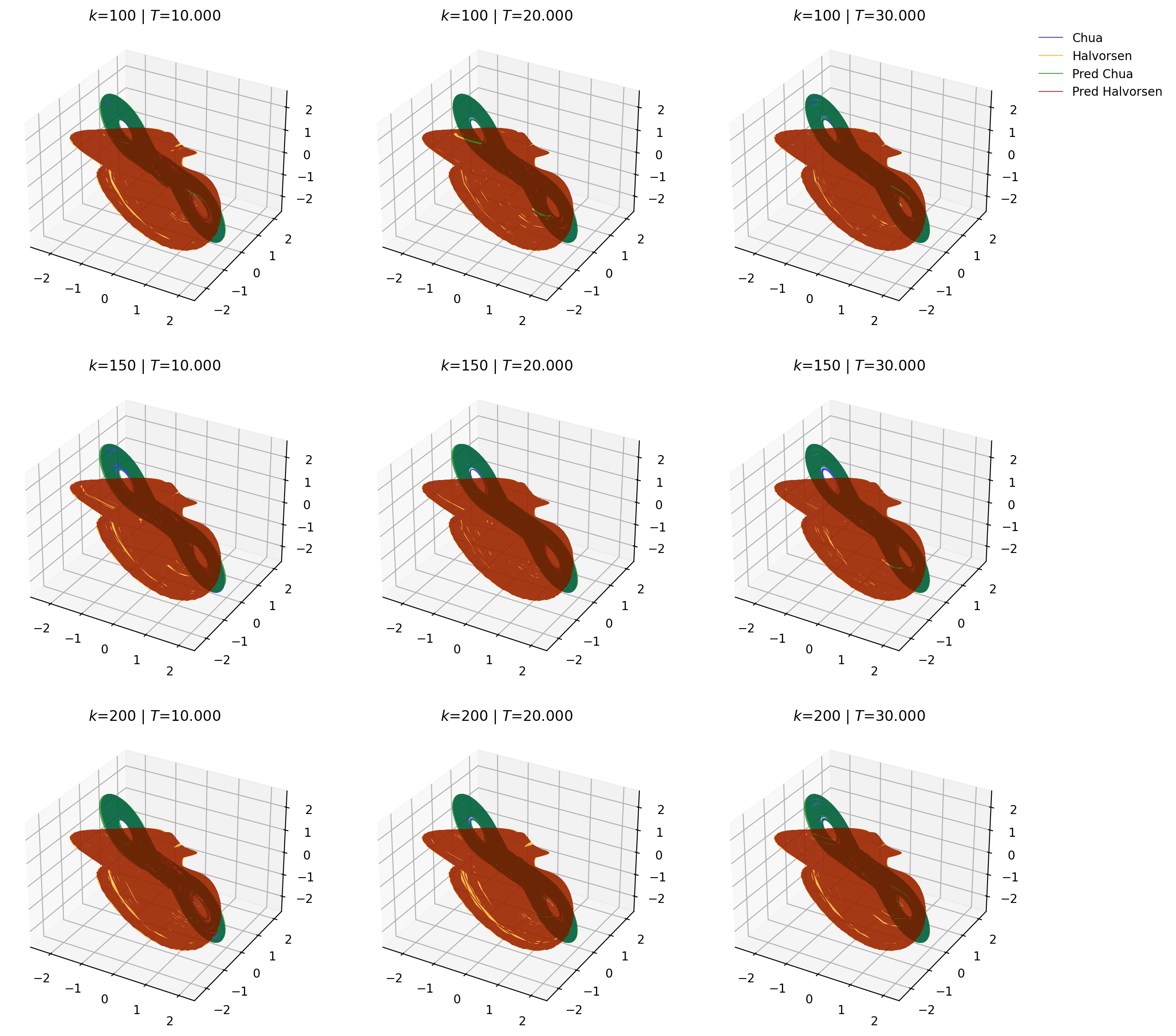}
    \caption{Predictions for the Chua and Halvorsen task for different $k$ and $T$ values.}
    \label{fig:Chua + Halvorsen predictions}
\end{figure}

% \newpage
\section*{Supplementary Note: Rucklidge and Windmi task}

For a final validation of our approach on the two attractor tasks we consider reconstructing chaotic attractors from the Rucklidge and Windmi systems both integrated with a time step size of $dt=0.08$. We normalize both obtained trajectories to zero mean and unit standard deviation and overlay them in state space. We apply the attractor reconstruction metrics over a prediction length $P=15{,}000$ across all tests.

\begin{figure}[t]
    \centering
    \includegraphics[width=\linewidth]{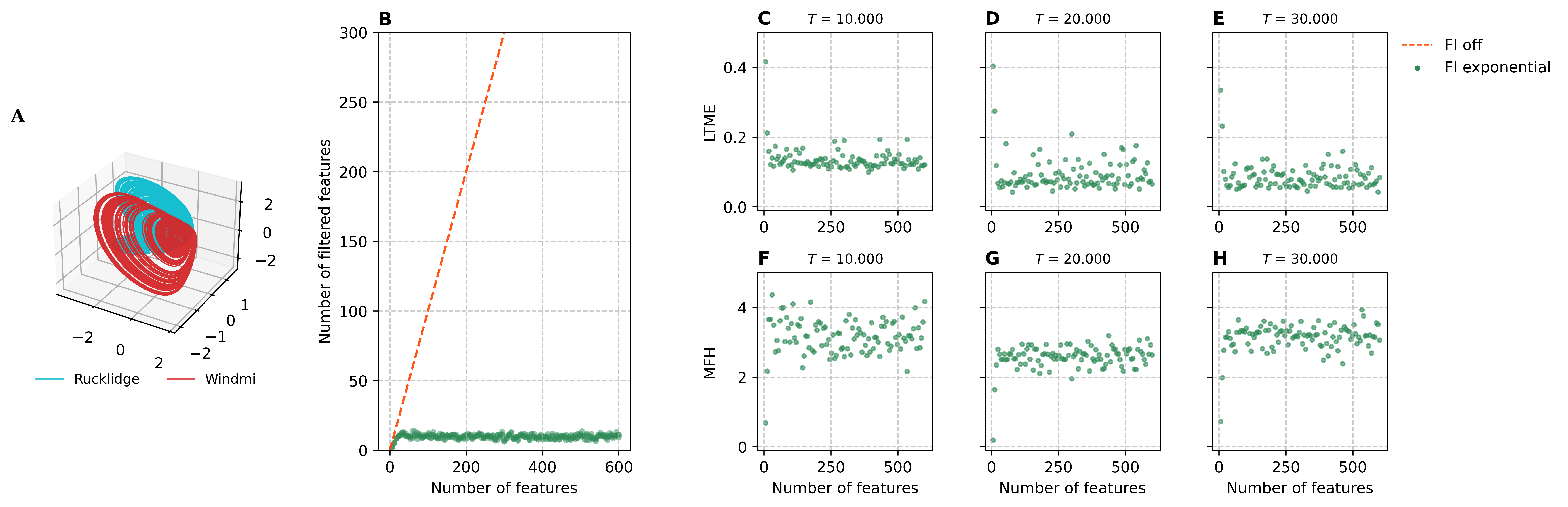}
    \caption{\textbf{Performance metrics of the exponential scheme on the Rucklidge and Windmi task}. Each data point represents the mean over five random initialized ERT instances. \textbf{A} Rucklidge and Windmi overlapping in state space, normalized to mean 0 and standard deviation 1, with scaling parameter $\gamma = 5$. \textbf{B} The number of input features versus the number of filtered features for the VAR feature model. There is no significant filtering for few features. The number of filtered features plateaus for a larger initial number of features with slightly higher variance observed than for the other two attractor tasks. The Long-term measurement error is displayed in \textbf{C}-\textbf{E} and the mean forecast horizon is displayed in \textbf{F}-\textbf{H}.
    }
    \label{fig:Rucklidge + Windmi combined}
\end{figure}
\begin{figure}[t]
    \centering
    \includegraphics[width=0.70\linewidth]{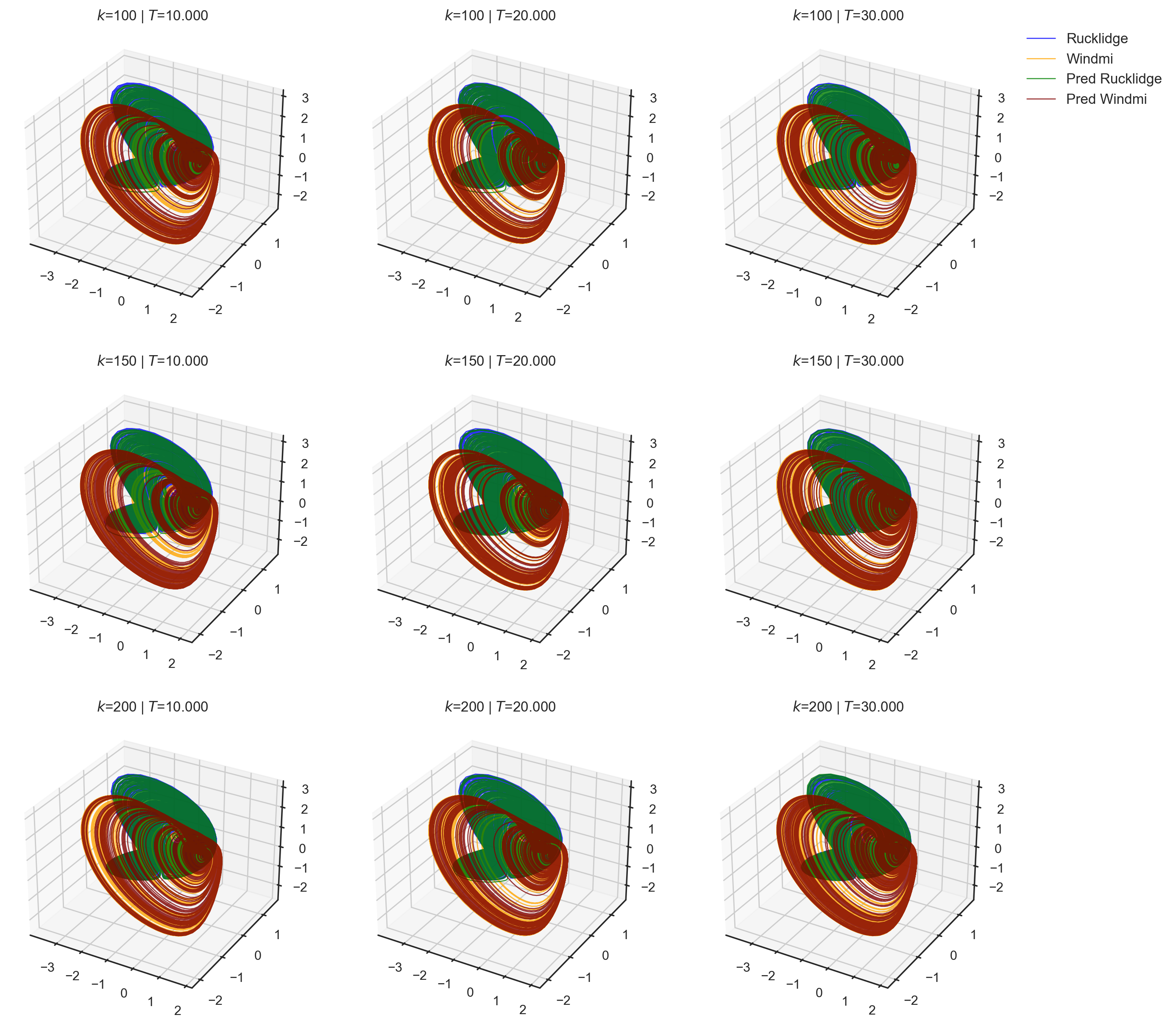}
    \caption{Predictions for the Rucklidge and Windmi task for different $k$ and $T$ values.}
    \label{fig:Rucklidge + Windmi predictions}
\end{figure}

% \newpage

\section*{Supplementary Note: Selected processing delays}

We analyze how selected memory delays influence long-term memory recall. Using the optimized NGRC+ERT system from the 4-attractor-task from the Results section, we shift the two most recent short-term memory features associated with each coordinate and simulate signal transmission lags by introducing time-delays of size $n$ to each selected feature.
We find that these defects lead to similar observations as for the simulated sensory delay, where the NGRC+ERT system loses its initially optimized functioning by recalling attractors it was never trained on.

\begin{figure}[t]
    \centering
    \includegraphics[width=\linewidth]{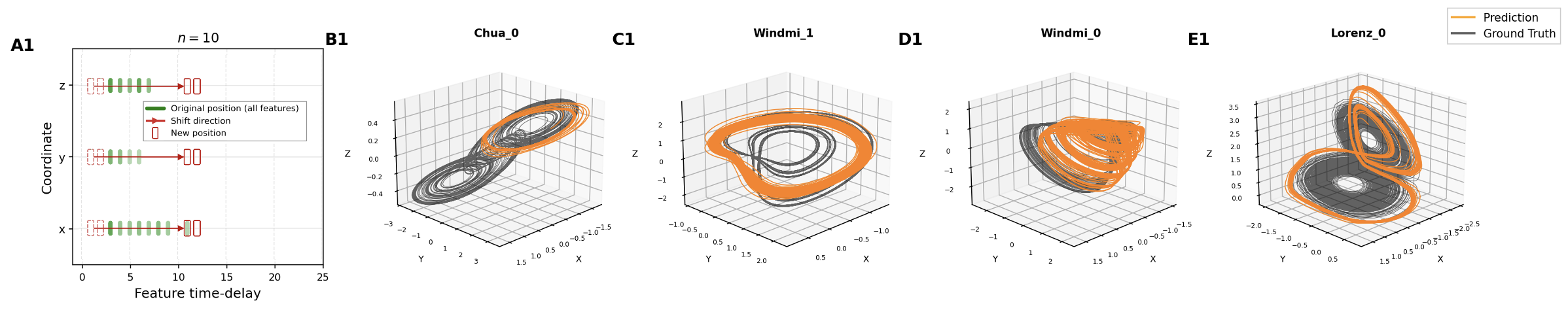}
    \vspace{0.5cm} % Adjust spacing between images as needed
    
    \includegraphics[width=\linewidth, trim=0 0 10px 0, clip]{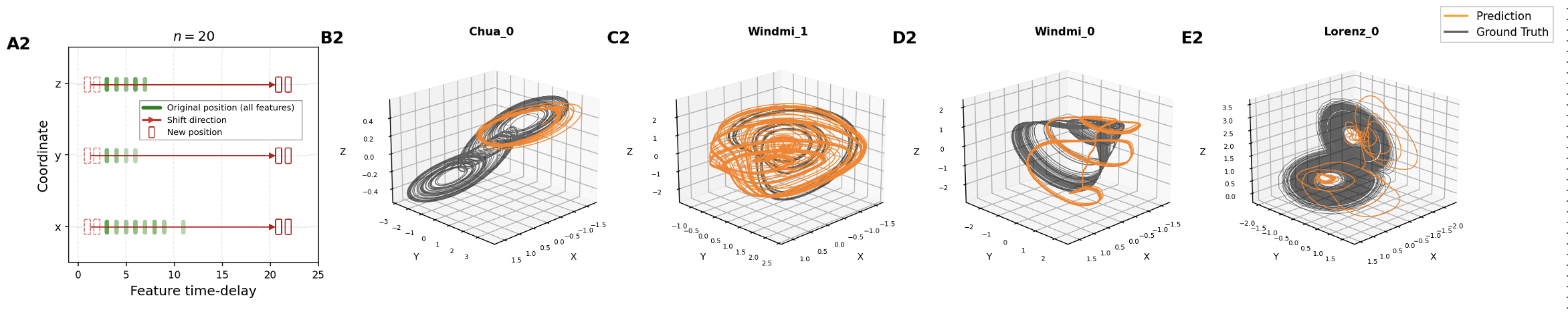}
    \caption{\textbf{Four attractor attractor task with processing time-delays in the two most recent memory features of each coordinate.} 
    \textbf{Top:} Lag 10 - Recalling the four attractors when defects in selected memory features are present provoke generated attractors. 
    \textbf{Bottom:} Lag 20 - When defects increased, changing behaviors of generated attractors emerge.}
    \label{fig:4attractorLagComparison_selcted}
\end{figure}

\begin{figure}[t]
    \centering
    \includegraphics[width=0.95\linewidth]{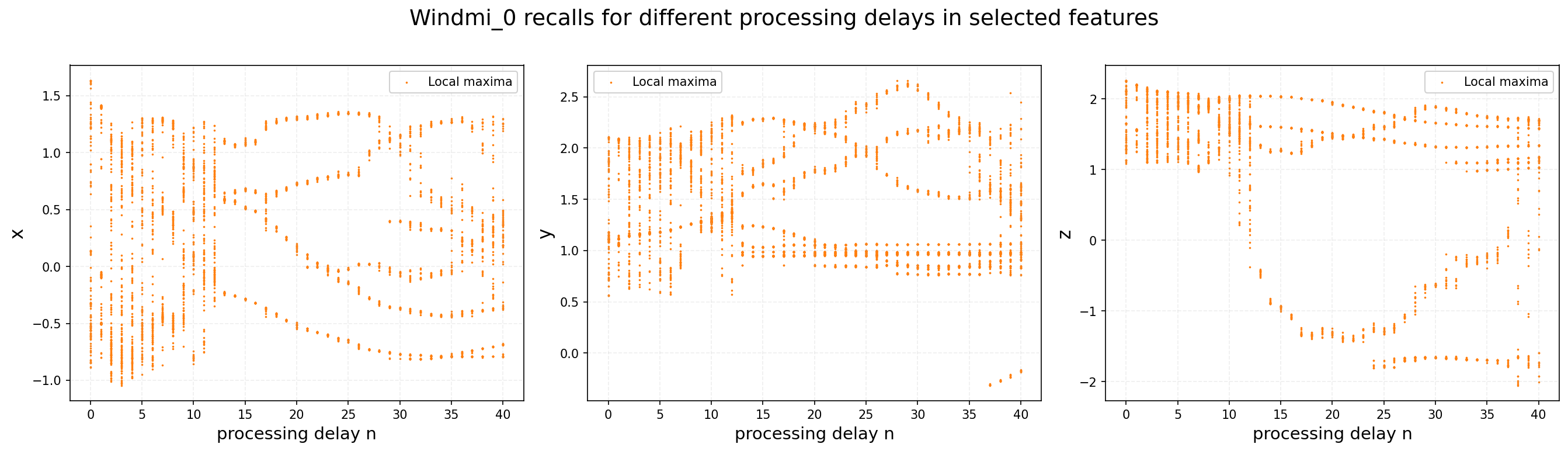}
    \caption{\textbf{Windmi$\_$0 recalls for processing delays in the two most recent features.}
    For Windmi$\_$0 attractor recalls when processing time-delays of the two most recent features of each coordinate, rich dynamical behaviors with emerging periodic and chaotic regimes the system was never trained on are observed.}
    \label{fig:4attractorBif_delay}
\end{figure}

\begin{figure}[t]
    \centering
    \includegraphics[width=0.95\linewidth]{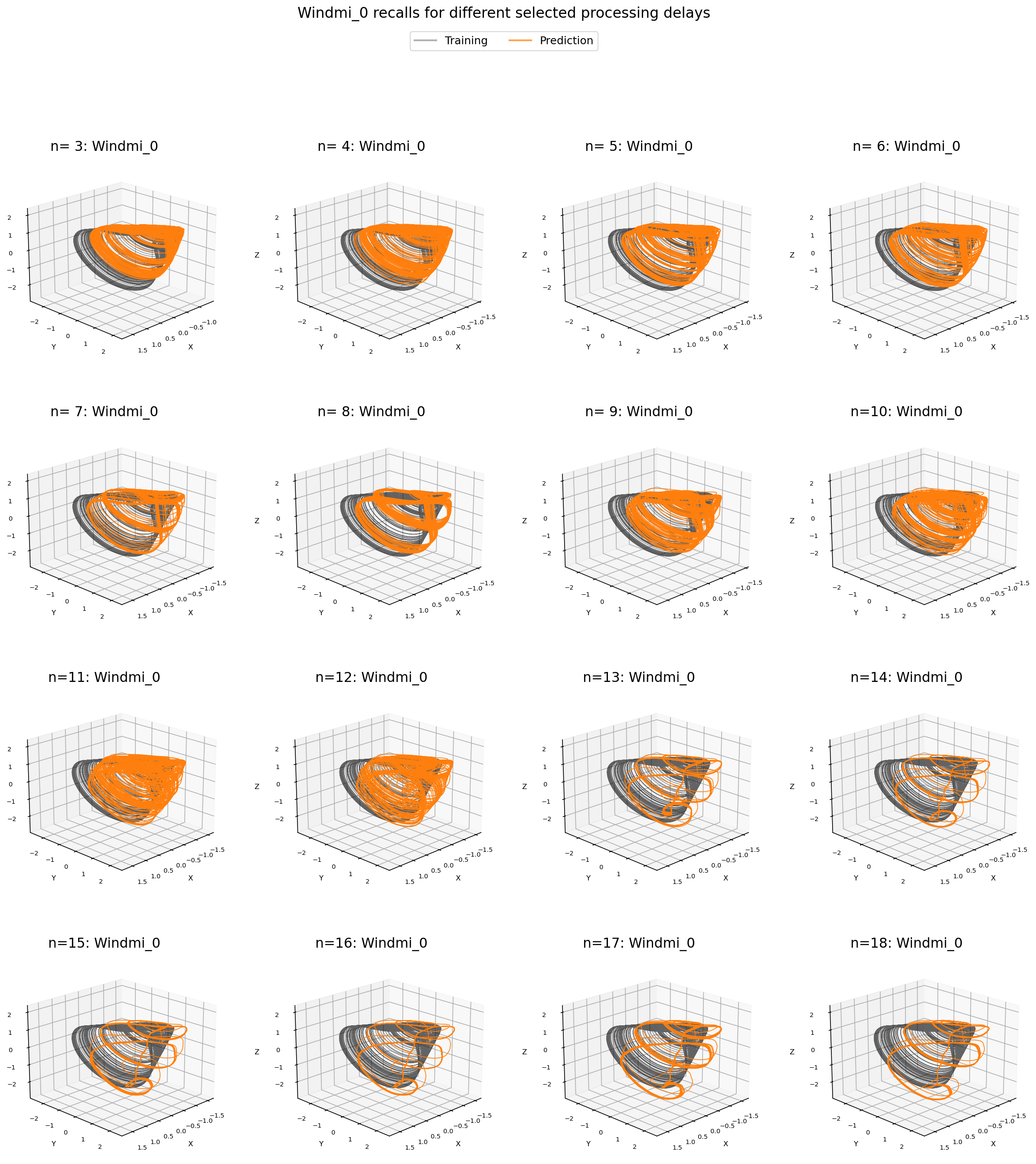}
    \caption{\textbf{Windmi$\_$0 recall dynamics for processing delays in the two most recent features.} Note sudden change from chaotic to quasi-periodic dynamics between $n=12$ and $13$.
    }
    \label{fig:windmi0_grid}
\end{figure}

\section*{Supplementary Note: Failure modes from defects in short-term memory processing}

\begin{figure}[t]
    \centering
    \includegraphics[width=0.95\linewidth]{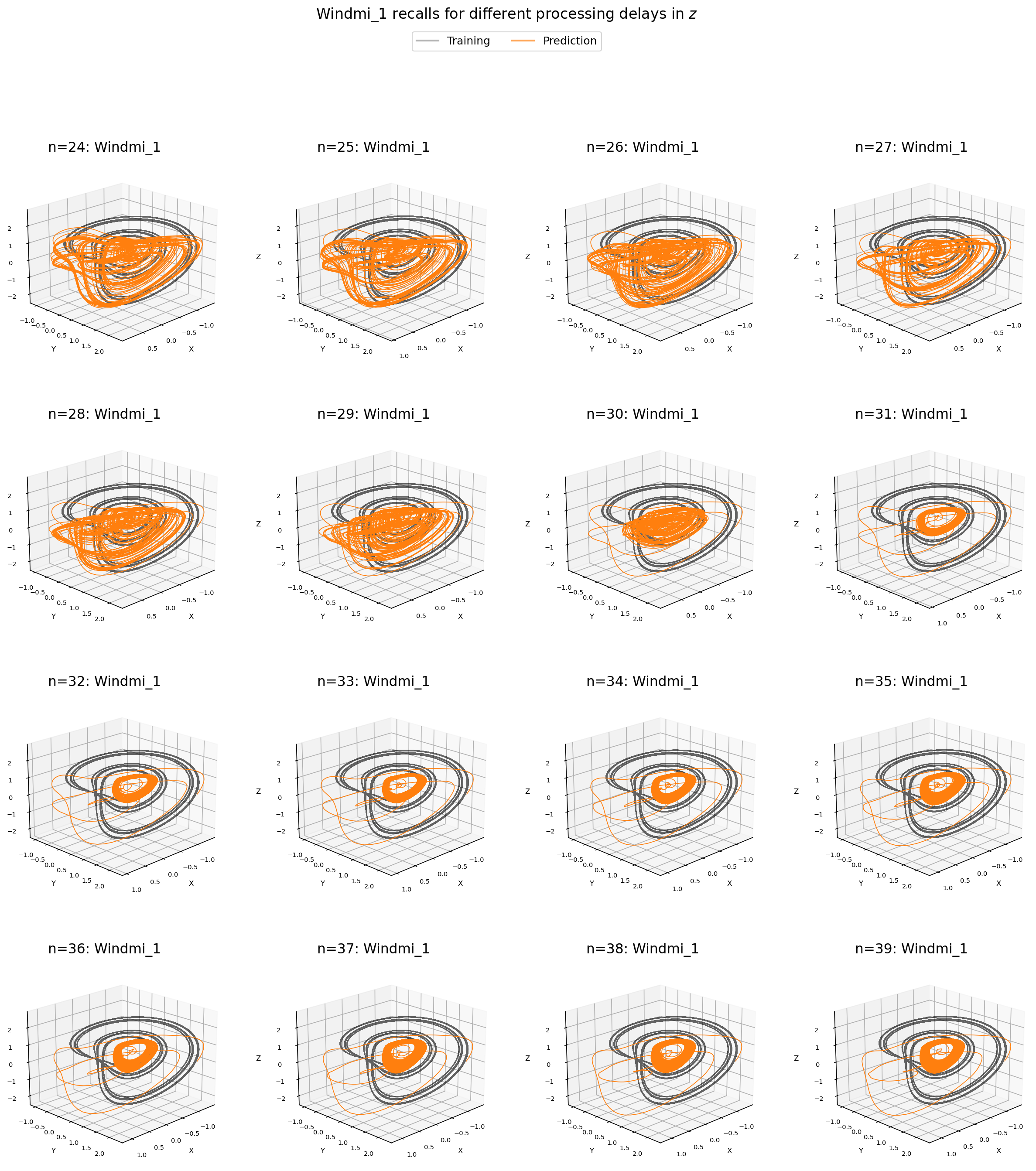}
    \caption{\textbf{Windmi$\_$1 recall dynamics for processing delays in $z$.} Note interior crisis taking place between $n=29$ and $30$, evidenced by sudden shrinking of attractor.
    }
    \label{fig:windmi1_grid}
\end{figure}

\end{document}